\documentclass[prb,showpacs,twocolumn]{revtex4-1}
\usepackage{graphicx,amssymb,amsmath,color}

\begin{document}

\newcommand{\supt}[1]{^{\text{#1}}}
\newcommand{\subt}[1]{_{\text{#1}}}
\newcommand{\Th}{T_{\text{H}}}
\newcommand{\TB}{T_{\text{B}}}
\newcommand{\Td}{T_{\text{d}}}
\newcommand{\subzz}{_{\text{zz}}}
\newcommand{\subac}{_{\text{ac}}}
\newcommand{\subim}{_{\text{im}}}
\newcommand{\supzz}{^2{\text{zz}}}
\newcommand{\supac}{^{\text{ac}}}
\newcommand{\supim}{^{\text{im}}}
\newcommand{\bsy}[1]{\boldsymbol{#1}}
\newcommand{\imi}{i}
\newcommand{\bra}[1]{\langle #1 |}
\newcommand{\ket}[1]{|#1\rangle}
\newcommand{\braket}[2]{\langle #1 |#2 \rangle}
\newcommand{\mypara}[1]{\paragraph{#1}~\\}
\newcommand{\sgn}{\text{sgn}}
\newcommand{\Gsc}{G^{\text{sc}}}
\newcommand{\sigmat}{\sigma_{\text{t}}}
\newcommand{\sigman}{\sigma_{\text{n}}}
\newcommand{\sigmaa}{\sigma_{\bsy{a}}}
\newcommand{\sigmab}{\sigma_{\bsy{b}}}
\newcommand{\tr}{\text{Tr}}
\newcommand{\taud}{\tau_{\text{d}}}
\newcommand{\taudexp}{{\displaystyle\tau_{\text{d}}}}
\newcommand{\thetal}{\theta_{\text{loop}}}
\newcommand{\bsig}{\bsy{\sigma}}
\newcommand{\dagg}{^{\dagger}}
\newcommand{\cdotsh}{\!\cdot\!}
\newcommand{\vF}{v_{\scriptscriptstyle F}}
\newcommand{\kE}{k_{\scriptscriptstyle E}}
\newcommand{\lambdaE}{\lambda_{\scriptscriptstyle E}}
\newcommand{\kmax}{k_\text{max}}
\newcommand{\sigmav}[1]{\sigma_{#1}}
\newcommand{\sDelta}{{\scriptstyle \Delta}}
\newcommand{\OpI}{\hat{\mathcal{I}}}
\newcommand{\I}{{\mathcal{I}}}
\newcommand{\OpIs}{\hat{\mathcal{I}}_{\text{s}}}
\newcommand{\Is}{{\mathcal{I}}_{\text{s}}}
\newcommand{\OpIl}{\hat{\mathcal{I}}_{\text{l}}}
\newcommand{\Il}{{\mathcal{I}}_{\text{l}}}
\newcommand{\intl}{\int\limits}
\newcommand{\unit}[1]{\,\text{#1}}
\newcommand{\rhoosc}{\rho_{\text{osc}}}
\newcommand{\rhobar}{\bar{\rho}}
\newcommand{\Ktilde}{{K}}
\newcommand{\Ktildes}{{K}^{\text{s}}}
\newcommand{\Ktildesd}{{K}^{\text{s}\dagger}}
\newcommand{\Ktildev}{{K}^{\text{v}}}
\newcommand{\Ktildevd}{{K}^{\text{v}\dagger}}
\newcommand{\disptauac}{{\displaystyle \tau_{\text{ac}}}}
\newcommand{\disptaud}{{\displaystyle \tau_{\text{d}}}}
\newcommand{\disptauB}{{\displaystyle \tau_{\text{B}}}}
\newcommand{\Wac}{W_{\text{ac}}}
\newcommand{\Tmin}{T_{\text{min}}(\varepsilon)}

\title{
Edge effects in graphene nanostructures:\\
    II. Semiclassical theory of spectral fluctuations and quantum transport \\
}

\author{J\"urgen Wurm}
\author{Klaus Richter}
\affiliation{Institut f\"ur Theoretische Physik, Universit\"at Regensburg, D-93040 Regensburg, Germany}

\author{ \.{I}nan\c{c} Adagideli}
\affiliation{Faculty of Engineering and Natural Sciences, Sabanc\i~ University, Orhanl\i~ - Tuzla, 34956, Turkey}

\date{\today}

\begin{abstract}

We investigate the effect of different edge types on the statistical properties of both the energy spectrum of closed graphene billiards and
the conductance of open graphene cavities in the semiclassical limit. To this end, we use the semiclassical Green's function for ballistic graphene flakes that we have derived in Reference \onlinecite{PartI}. 
First we study the spectral two point correlation function, or more precisely its Fourier transform the spectral form factor, starting
from the graphene version of Gutzwiller's trace formula for the oscillating part of the density of states. We calculate the two leading order contributions  to the spectral form factor, paying particular attention to the influence of the edge characteristics of the system.
Then we consider transport properties of open graphene cavities. We derive generic analytical expressions for the classical conductance, the weak localization correction, the size of the universal conductance fluctuations and the shot noise power of a ballistic graphene cavity. Again we focus on the effects of the edge structure.
For both, the conductance and the spectral form factor, we find that edge induced pseudospin interference affects the results significantly. In particular intervalley coupling mediated through scattering from armchair edges is the key mechanism that governs the coherent quantum interference effects in ballistic graphene cavities.

\end{abstract}


\maketitle

\section{Introduction}
\label{sec:intro}

Nearly twenty-five years ago, Berry and Mondragon \cite{Berry1987} introduced the `neutrino billiard', 
a Dirac Hamiltonian describing a massless spin one-half particle in a plane with lateral confinement, as a fictitious, conceptually
simple quantum system which exhibits time-reversal symmetry (TRS) breaking without magnetic fields. 
They demonstrated that the energy eigenvalues of such a Dirac billiard are statistically distributed 
according to the corresponding Gaussian unitary ensemble (GUE) of random matrix theory (RMT), i.\,e.\,the ensemble of random hermitian matrices without TRS,
if the quantum system possesses a classical counterpart that is chaotic. 

Ballistic cavities built from monolayer graphene (for recent reviews see Refs.\,\onlinecite{Peres2010, Abergel2010, Das2011}) with Fermi energy close to the 
Dirac point, have been proposed as realizations of such a neutrino billiard\cite{Ponomarenko2008, Guttinger2008, Guttinger2010}.
In fact due to the coexistence of the Dirac points associated with the two independent valleys  
in the graphene band structure, graphene-based billiards represent two copies of a Dirac billiard mutually
coupled through \emph{intervalley scattering}. For vanishing intervalley coupling, the entire graphene Hamiltonian
is composed of two decoupled sectors with degenerate sets of eigenvalues each set obeying GUE statistics just like the neutrino billiard. 
However, for sufficiently strong intervalley coupling, one expects an eigenvalue statistics according to the Gaussian 
orthogonal ensemble (GOE), because the entire graphene Hamiltonian is time-reversal invariant, i.\,e.\,it is a real symmetric matrix
in the absence of additional magnetic fields. Therefore intervalley coupling 
is an effective symmetry restoring parameter that allows tuning from GUE to GOE behavior
in graphene.

In disordered graphene samples, short ranged scattering potentials or vacancies cause intervalley scattering.
Depending on the corresponding intervalley scattering time and other relevant inverse scattering rates of the system (e.\,g.\,the phase coherence time),
the magnetoconductance of disordered bulk graphene is expected to show signatures of weak localization or weak antilocalization\cite{McCann2006}. This theoretical picture has been confirmed experimentally to some extent by the measurement of weak localization\cite{Morozov2006, Tikhonenko2008, Berezovsky2010} and weak antilocalization\cite{Wu2007} in disordered graphene.

In contrast, in ballistic graphene nanostructures (with elastic mean free paths on scales of several microns\cite{Mayorov2011}), it is the edges that are responsible for intervalley scattering. However, not all types of edges are effective intervalley scatterers: While scattering of quasiparticles from a zigzag or an infinite mass type edge is an \emph{intravalley} scattering process, armchair edges couple quasiparticle states from both Dirac points. Recent experiments on nanostructured graphene\cite{Eroms2009} have revealed that the magnetoconductance in such samples cannot be consistently described using the bulk theory of Ref.\,\onlinecite{McCann2006} and signatures of ballistic transport were found to be non-negligible. While several numerical studies have investigated quantum interference effects in spectral and transport properties of ballistic graphene nanostructures\cite{Wurm2009, Libisch2009, Bardarson2009, Yang2011}, so far there is no analytical theory that is capable of treating quantum interference effects in  arbitrarily shaped graphene nanostructures. In this work, we provide a semiclassical theory for the conductance of ballistic graphene structures as well as for spectral correlations, particularly focusing on the effects of different edge types.
Our semiclassical approach requires $L \gg \lambdaE$ with $\lambdaE$ the Fermi wavelength
and $L$ the system size, while at the same time the energies should be compatible
with the assumption of a linear dispersion law in the effective Dirac theory for graphene, i.\,e.\,$\lambdaE \ll a$, with the graphene lattice constant $a$.
Fortunately, both conditions are well fulfilled for most mesoscopic graphene structures:
For example, for a system size of order $L \sim $ 100 nm, there are thousands of levels in the linear dispersion regime, where the Dirac equation is valid.

This is the second paper of a two-paper series on edge effects in graphene
nanostructures. In the first paper \cite{PartI}, referred to as paper I in the following, 
we derived an expression for the single-particle Green's function of a ballistic or weakly disordered graphene 
structure in terms of multiple reflections from the system boundary. Using this expansion, we could
account for different types of boundary conditions (zigzag, armchair, infinite mass 
or combinations of those), while the propagation inside the system was treated on the
level of the effective Dirac theory. Furthermore, we obtained expressions for the mean density of states (DOS) valid in the semiclassical limit, 
highlighting the role of edge states at
zigzag boundary segments, as well as for the oscillating part of the DOS in terms of
periodic orbits for classically chaotic and certain integrable graphene cavities.
Graphene edge phenomena and the effects from sublattice and valley pseudospin dynamics 
enter into those trace formulae through (traces over) pseudospin propagators evolving
along the classical orbits that otherwise are the same as in a corresponding
non-relativistic billiard.

Here we consider the spectral statistics of closed chaotic graphene cavities
as well as the conductance of open graphene cavities coupled to (two) leads. We address the question as to how their universal properties are governed by the underlying Dirac-type 
dynamics together with the graphene-specific edge effects. In particular, we 
study signatures of the afore mentioned crossover, mediated through intervalley scattering, between unitary and orthogonal symmetry classes
in spectral and transport observables. 
We are thereby able to give semiclassical explanations and analytical expressions for results obtained earlier in Ref.\,\onlinecite{Wurm2009} mainly numerically and in terms of symmetry arguments. There it was shown that intervalley coupling (mediated through scattering at armchair edges) acts differently on spectral and transport properties of classically chaotic graphene cavities: While in quantum transport a crossover from unitary to orthogonal behavior should be observable with increasing intervalley coupling, spectral statistics probing the scale of the mean level spacing is predominantly governed by the orthogonal symmetry class, even for rather weak armchair scattering.
Here we study the spectral statistics at energy scales larger than the mean level spacing, where the unitary symmetry should be observable even at moderate intervalley scattering.

To this end we start from the semiclassical expressions 
for the Green's function and DOS in terms of interfering classical trajectories, derived in paper I, which
we introduce in Sec.\,\ref{sec:Green-symmetry}. 
In Sec.\,\ref{sec:spec-stat} we address spectral fluctuations, more specifically we derive 
semiclassical expressions for the spectral form factor, the Fourier transform of the spectral 
two-point correlator characterizing spectral statistics. We find that the leading order and the next to leading order contributions are strongly influenced 
by the edges. The total amount of armchair edges is the relevant quantity that determines the size of graphene specific correlations: 
Our results suggest that evidence for partial (effective) breaking of the time reversal symmetry (TRS) is visible in the spectral correlations
at scales of many mean level spacings. The reader with main interest in transport can also skip Sec.\,\ref{sec:spec-stat} and directly go to Sec.\,\ref{sec:transport}, which is devoted to mesoscopic phenomena in coherent transport. We study imprints of the edges and the chaotic dynamics in confined graphene systems on weak localization, conductance 
fluctuations, and shot noise. 

We start from an exact expression for the Green's function in terms of multiple scatterings from the boundary\cite{Balian1970, Hansson1983, Adagideli2002, PartI} and generalize advanced semiclassical 
techniques, introduced in Refs.\,\onlinecite{Sieber2001, Richter2002}, for graphene, accounting for classical trajectory correlations, which have proven 
essential for understanding spectral statistics and quantum phenomena in ballistic transport through chaotic 
Schr\"odinger billiards.
In this way we derive analytical expressions for the observables mentioned above that allow us to predict the edge dependence of several
quantities in the semiclassical regime. While the details of the edge structure are crucial for the form factor, the weak localization correction and the conductance fluctuations, it turns out that shot noise
is not affected to leading order in the inverse channel numbers of the leads.
We finally conclude and give an outlook in Sec.\,\ref{sec:conclusions}, and 
collect longer derivations in three appendices.

\section{Green's function and antiunitary symmetries}
\label{sec:Green-symmetry}

\subsection{Semiclassical Green's function for graphene billiards}
\label{ssec:semiclassics}

In  Secs.~\ref{sec:spec-stat} and \ref{sec:transport} we employ the Green's function in order 
to study correlations in the spectra and the conductance of graphene flakes in the semiclassical limit.
In paper I we derived an exact expression for the Green's function of a graphene flake in terms of 
a multiple reflection expansion, where each term in this expansion corresponds to the sum of Feynman paths 
with a specific number of reflections at the system boundary. In this section we repeat our main results that will be useful for the present paper.

\begin{figure}
 \centering
 \includegraphics[width=0.4\textwidth]{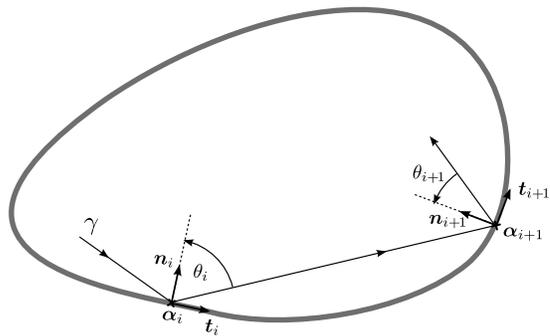}
\caption{\small{Scheme of two classical reflections along an orbit $\gamma$. The normal directions at the reflection points $\bsy{\alpha}_i$ and $\bsy{\alpha}_{i+1}$ are denoted by unit vectors $\bsy{n}_i$ and $\bsy{n}_{i+1}$, respectively. The tangential directions are $\bsy{t}_i = \bsy{n}_i \times \hat{\bsy{z}}$ and $\bsy{t}_{i+1} = \bsy{n}_{i+1} \times \hat{\bsy{z}}$, while the reflection angles are $\theta_i$ and $\theta_{i+1}$, respectively.
}}
\label{Fig:refangle}
\end{figure}

For a confined ballistic graphene structure quasiparticle dynamics is described by the effective Dirac Hamiltonian 
\begin{equation}
\label{eq:Ham}
 H = \vF \tau_0 \otimes \bsy{\sigma}\cdot \bsy{p}\,,
\end{equation}
with appropriate boundary conditions at the edges. Here the $\{\sigma_i\}$ denote Pauli matrices in sublattice pseudospin space and Pauli matrices in valley-spin space are represented by  $\{\tau_i\}$, while $\sigma_0$ and $\tau_0$ are unit matrices acting on the corresponding spin space. Then the full Green's function is given in the semiclassical limit by (cf. Eq.~(78) in I)
\begin{equation}
\label{eq:Gsc}
 G(\bsy{x}, \bsy{x'}) 
 = \frac{\vF}{\sqrt{8\pi \hbar}} \!\!\sum_{\gamma(\bsy{x},\bsy{x'})} \!D_{\gamma}
\,K_{\gamma} e^{i S_{\gamma}/\hbar+i\mu_{\gamma}\pi/2}  \, .
\end{equation}
Here the sum runs over all \textit{classical orbits} $\gamma$ leading from point 
$\bsy{x'}$ to $\bsy{x}$.
These classical orbits consist of straight lines that are connected by specular reflections at the boundary,
so that the condition of least action is satisfied. This situation is similar to billiards with spin orbit interaction (SOI) where the entire effect of the SOI is contained in the spin matrix with the same trajectories\cite{Mathur1992, Zaitsev2005, Adagideli2010}.
In Eq.\,(\ref{eq:Gsc}) $S_{\gamma}= \hbar \kE L_{\gamma}$ and $\mu_{\gamma}$ are the classical action ($\kE$ and $L_{\gamma}$ are the Fermi momentum and the length, respectively) and the number of conjugate points for the classical orbit $\gamma$, and 
\begin{equation}
 D_\gamma = 
\frac{1}{\vF}\left|\frac{\partial{x}_{\perp}}{\partial {p'}_\perp}\right|^{-1/2}_{\gamma}\,
\end{equation}
denotes an element of the stability matrix of the path $\gamma$. Here $p'_\perp$  and $x_\perp$ are the components of the initial momentum  and final position 
perpendicular to the trajectory.
In Eq.~(\ref{eq:Gsc}), 
\begin{eqnarray}
\label{eq:pspinprop}
 K_\gamma &=& \prod_{i= N_\gamma}^{1} K_i \left(1+\sigmav{\bsy{\alpha}_{1},\bsy{x'}}\right)
\end{eqnarray}
is the pseudospin propagator along the orbit $\gamma$, with $\sigmav{\bsy{\alpha}_{1},\bsy{x'}} = \bsy{\sigma}\cdot(\bsy{\alpha}_{1} - \bsy{x'})$. The product in Eq.\,(\ref{eq:pspinprop}) runs over all $N_\gamma$ reflections that occur at boundary points $\bsy{\alpha}_i$ along the orbit $\gamma$, with the edge dependent pseudospin rotations for reflections at zigzag (zz), armchair (ac), and infinite mass (im) type edges
\begin{eqnarray}
\label{eq:K_i}
 K_i = \pm i
\left\{
\begin{array}{ll}
 e^{\pm i\theta_i\tau_z}\tau_z\otimes\sigmav{\bsy{t}_i} & \text{for~zz}\,, \\
 \tau_ye^{i2Kx_i\tau_z}\otimes e^{i\theta_i \sigma_z}\sigma_z & \text{for~ac}\,,\\
 \tau_z\otimes e^{i\theta_i \sigma_z}\sigma_z & \text{for~im}\,.
\end{array}
\right.
\end{eqnarray}
Here $\theta_i$ denotes the reflection angle at the $i$-th bounce with the boundary, while
$\bsy{t}_i$ and $x_i$ are the direction of the tangent to the boundary and the
$x$-coordinate of the reflection point $\bsy{\alpha}_i$, respectively (see Fig.\,\ref{Fig:refangle}). Further we define for a given vector $\bsy{v}$ the Pauli matrix $\sigma_{\bsy{v}} =\bsy{\sigma} \cdot \bsy{v}$. $K=4\pi/3a$ is the distance of the Dirac points from the $\Gamma$-point 
of the Brillouin zone. For reflections at infinite mass type edges, the sign in Eq.\,(\ref{eq:K_i}) is equal to the sign of the mass potential outside of the system\cite{Berry1987}. 
For zigzag edges the sign is determined by the type of sublattice at the zigzag edge. For an A-edge, the upper sign is valid and for a B-edge the lower sign. For armchair edges, the upper sign is valid when the order of the atoms within each dimer is $A$-$B$ along the direction
of $\bsy{t}_i$, and the lower sign is valid for $B$-$A$ ordering. 

Equations (\ref{eq:Gsc} - \ref{eq:K_i}) specify the contribution of a given 
classical orbit $\gamma$ to the semiclassical Green's function.
Note that the dynamical part is identical to the semiclassical Green's function of 
a Schr\"odinger system, while it is the  pseudospin propagator $K_\gamma$, through which the graphene specific physics enters. This relation allows us to use many results from the semiclassical theory of Schr\"odinger billiards. 
\subsection{Antiunitary symmetries of the effective theory}

In order to compare the results from our semiclassical theory with universal RMT predictions, we need to 
consider the relevant antiunitary symmetries of the Hamiltonian $H$, Eq.\,(\ref{eq:Ham}). An antiunitary operator is given by the complex conjugation followed by 
a unitary operator. We assume that all spatial symmetries, such as invariance under reflection or rotation, are broken. Thus the only relevant unitary operations are rotations in the pseudospin spaces. As mentioned above we adapt here the valley-isotropic basis, in which the bulk Hamiltonian is invariant under the transformations $\mathcal{T}_i^{-1} H \mathcal{T}_i$ with\cite{Note1}
\begin{equation}
 \mathcal{T}_i = \sigma_y \otimes \tau_i\,\mathcal{C}\,, \quad i\in \{0,x,y\}\,.
\end{equation}
Furthermore we note that $\mathcal{T}_y^2=1$ while \mbox{$\mathcal{T}_0^2=\mathcal{T}_x^2=-1$}.  That means $\mathcal{T}_y$, the overall TRS that
connects states from opposite valleys, drives the system into the orthogonal symmetry class.
This symmetry is not broken by any of the boundaries we consider. However, we will see that it is only relevant if intervalley scattering is
present. In the absence of intervalley scattering, the effective intravalley TRS $\mathcal{T}_0$ gives rise to the symplectic universality class.
However, for the boundary conditions that do not couple the valleys (infinite mass, zigzag), $\mathcal{T}_0$ is not preserved, and thus it is irrelevant for
ballistic systems. 
Finally $\mathcal{T}_x$ is also a symplectic symmetry. It is broken by intervalley scattering armchair edges and left intact by zigzag or infinite mass type edges.
For preserved valleys it thus ensures the (Kramers) degeneracy of the two valleys.

We note that, since the bulk Hamiltonian has a particle-hole symmetry ($ \sigma_x  H^* \sigma_x  = -H$), the chiral universality classes are relevant at zero energy. In this paper, however, we consider Fermi energies away from zero, where the normal and the chiral universality classes lead to the same results\cite{Gnutzmann2004}.

Sections \ref{sec:spec-stat} on spectral statistics and Sec.\,\ref{sec:transport} on transport can both be read independently. We begin with the analysis of spectral statistics which is conceptually slightly simpler.

\section{Spectral form factor of chaotic graphene billiards}
\label{sec:spec-stat} 

\subsection{Spectral form factor: \\
Definition and random matrix results}
Quite generally, the DOS can be split into a smooth part, i.\,e.\,the mean DOS $\bar{\rho}$, and an oscillating part $\rho_{\text{osc}}$,
\begin{equation}
 \rho(\kE) = \bar{\rho}(\kE) + \rho_{\text{osc}}(\kE)\,.
\end{equation}
While $\bar{\rho}$ contains information about the coarse structure of the spectrum, details
about the level distribution are in $\rho_{\text{osc}}$. The mean DOS $\bar{\rho}$ and the trace formulae for the oscillating DOS $\rho_{\text{osc}}$ were addressed in I. Here we use the trace formula for $\rho_{\text{osc}}$ to study statistical properties of the spectral fluctuations.
To this end, we focus on the spectral two-point correlator 
\begin{equation}
\label{eq:Rdef}
 R(\eta) = \frac{1}{\bar{\rho}(\kE)}\langle \rhoosc(\kE + \eta/2) \rhoosc(\kE - \eta/2) \rangle_{\kE}\,,
\end{equation}
where $\langle ~\rangle_{\kE}$ denotes averaging over a (classically) small window of the Fermi energy $\hbar \vF \kE$ or, respectively, on its Fourier transform the spectral form factor
\begin{equation}
\label{eq:Fdef}
 F(t) = \intl_{-\infty}^{\infty} \! d\eta\,R(\eta) e^{i 2\pi \eta t \rhobar(\kE)}\,.
\end{equation}
Here $t$ denotes the time in units of the Heisenberg time $\Th = 2\pi\bar{\rho}(\kE)/\vF$.
In the limit $t\rightarrow 0$, RMT predicts
the universal expression\cite{Haake2004}
\begin{equation}
\label{eq:FRMT2}
 F(t) =
\left\{
\begin{array}{cl}
 2t - 2t^2 + \mathcal{O}(t^3)& \text{for~GOE}\,, \\
 t & \text{for~GUE}\,,
 \\
\frac{1}{2}t +\frac{1}{4}t^2 + \mathcal{O}(t^3)& \text{for~GSE}\,,
\end{array}
\right. 
\end{equation}
which is expected to be valid for chaotic systems.
Systems with a TRS $\mathcal{T}$ and $\mathcal{T}^2=1$ (e.g. a free electron without magnetic field) belong to the orthogonal
symmetry class and systems without TRS (e.g. an electron in a magnetic field) to the unitary class.
Systems with a TRS $\mathcal{T}$ and $\mathcal{T}^2=-1$ (e.g. a spin 1/2 particle with broken spin rotational symmetry) are members of the 
symplectic class. 
However, as mentioned above, for ballistic graphene flakes the edges break the symmetry $\mathcal{T}_0$ that would lead to the symplectic universality class.
In the presence of spin-orbit interaction, which we neglect in this paper, the symplectic class could be relevant.

\subsection{The spectral form factor: \\Semiclassical evaluation }

\subsubsection{Gutzwiller trace formula for graphene}

For chaotic Schr\"odinger billiards, it is possible\cite{Berry1985, Sieber2001, Heusler2007} to connect the properties of classical orbits to $F(t)$. Using this relation the universal RMT form factor (\ref{eq:FRMT2}) has been derived semiclassically using 
Gutzwiller's trace formula\cite{Gutzwiller1990} for Schr\"odinger systems.
In the following we will extend this semiclassical theory to describe the spectral correlations of chaotic graphene 
systems. 

We start by
the following extension of Gutzwiller's trace formula (see Eq.\,(103) in I) to graphene:
\begin{equation}
\label{eq:Gutzwiller}
  \rho_{\text{osc}}(\kE) = \frac{\vF}{2\pi}\mathrm{Re}\sum_{\gamma} 
    A_\gamma \tr({K}_\gamma) \, e^{i\kE L_\gamma}\,.
\end{equation}
Here the sum runs over infinitely many \textit{periodic} classical orbits $\gamma$. 
The classical amplitudes $A_\gamma$ depend on the period, the stability and the number of conjugated
points of the corresponding orbit\cite{Gutzwiller1990}. As mentioned above, the trace over the pseudospin propagator $K_\gamma$ accounts for the graphene features. 

From Eqs.\,(\ref{eq:pspinprop},\,\ref{eq:K_i}) we know 
that the valley part of $K_\gamma$ contains an off-diagonal factor
\begin{equation}
 \tau_{K_i} =  \tau_y\, e^{i2Kx_i \tau_z}
\end{equation}
for every reflection at an armchair edge. This implies that the trace vanishes, $\tr\,K_\gamma = 0$, unless the total number of ac reflections $N_{\text{ac}}$ along $\gamma$ is even. If $N_{\text{ac}}$ is even, the pseudospin factor we have for an orbit with a total number of $N=N_\gamma$ reflections is (see App.\,\ref{app:pspintrace_1})
\begin{eqnarray}
\label{eq:trace-K}
 \tr\,K_\gamma = && \hspace*{-0.2cm} 4 f_\gamma \cos \left(\theta_\gamma+ N \pi/2\right) \nonumber \\
&\times& \cos\left(2K\Lambda_\gamma + \vartheta_\gamma + N\pi/2 \right)\,,
\end{eqnarray}
where $f_\gamma$ is a phase factor that depends on the exact sequence of ac, zz, and im type reflections,
\begin{equation}
\label{eq:refang}
 \theta_\gamma = \sum_{i=1}^{N} \theta_i
\end{equation}
is the sum over all reflection angles $\theta_i$ along the orbit $\gamma$,
\begin{equation}
\label{eq:Lambda_def}
 \Lambda_\gamma = \sum_{i=1}^{N_{\text{ac}}/2} \left(x_{{2i-1}} - x_{{2i}}\right)
\end{equation}
measures the differences in the $x$-coordinate of pairs of subsequent ac reflection points, and 
\begin{equation}
\label{eq:vartheta_def}
 \vartheta_\gamma =\sum_{i=1}^{N_{\text{zz}}} (-1)^{\text{s}_i}\vartheta_i\,.
\end{equation}
In Eq.\,(\ref{eq:vartheta_def}) the sum is restricted to the $N\subzz$ reflections at zz edges along $\gamma$, and $s_i$ is the number of ac reflections that occur \emph{after} the specific zz reflection $i$. Furthermore $\vartheta_i = +\theta_i$ for reflections at $A$-edges and $\vartheta_i = -\theta_i$ for reflections at $B$-edges.
We note that $\tr\,K_\gamma =  \tr\,K_{\gamma^{-1}}$ where $\gamma^{-1}$ and $\gamma$ are time reversed partners in the classical sense, i.\,e.\,they are identical up to the direction of movement.

\subsubsection{Products over phase-carrying paths}
\begin{figure}
 \centering
 \includegraphics[width=0.35\textwidth]{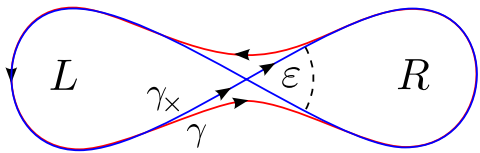}
\caption{\small{Scheme of the orbit pairs that contribute to the spectral form factor, Eq.\,(\ref{eq:Fsc}), to leading order beyond the diagonal approximation.
The orbit $\gamma_\times$ intersects itself with an angle $\varepsilon$, while the orbit $\gamma$ avoids this crossing.
Apart from this encounter region, the orbits follow each other exponentially closely, with the same propagation direction in the left loop $L$ and
opposite direction in the right loop $R$. Note that the real trajectories in a billiard consist of a series of many straight pieces and 
classical reflections at the system boundary between them, as shown in Fig.\,\ref{Fig:refangle}.}}
\label{Fig:figure8}
\end{figure}
We now evaluate the spectral form factor for graphene semiclassically. We insert Eq.\,(\ref{eq:Gutzwiller}) into the definition of $F(t)$ and obtain the approximate expression
\begin{equation}
\label{eq:Fsc}
  F(t) \approx \left<\sum_{\gamma, \gamma'} 
\frac{A_\gamma A_{\gamma'}^*Z_{\gamma,\gamma'}}{4\Th}
e^{i\delta S_{\gamma,\gamma'}/\hbar} \delta\left(T\! - \frac{T_\gamma\!+T_{\gamma'}}{2}\right)\right>_{\kE} \hspace*{-0.3cm},
\end{equation}
where $T_\gamma = L_\gamma /\vF$ is the period of an orbit, $T= t\, \Th$, 
and the short notation for the pseudospin traces
\begin{equation}
\label{eq:defZ}
 Z_{\gamma,\gamma'} = \tr\,\Ktilde_\gamma\tr \,\Ktilde_{\gamma'}^\dagger\,.
\end{equation}
The action difference between the orbits is given by $ \delta S_{\gamma,\gamma'} = \hbar\kE (L_\gamma-L_{\gamma'})$.

If $(\gamma, \gamma')$ is an arbitrary, uncorrelated pair of orbits, $e^{i\delta S_{\gamma,\gamma'}/\hbar}$ is generically a rapidly
oscillating function of $\kE$ in the semiclassical limit. Thus after $\kE$ averaging, the corresponding contribution to $F$ vanishes and the relevant contributions to $F$ come from classes of orbit pairs with small or even vanishing action difference.
That means the actions of the orbits have to be classically correlated.
The most obvious class of correlated orbit pairs 
are captured by the so-called diagonal approximation, where only equal action pairs $\gamma' = \gamma$ (and $\gamma' = \gamma^{-1}$) are considered. These orbit pairs give rise to the leading order ($\sim t$) terms in 
the expansion (\ref{eq:FRMT2}) for $F(t)$ \cite{Berry1985}. The second order terms ($\sim t^2$) in the GOE case are connected to another class of periodic orbit pairs: loop contributions that are sketched in Fig.\,\ref{Fig:figure8}. These two orbits follow each other closely for most of the time but one of them has a self-crossing 
($\gamma_\times$) while the other one \textquoteleft avoids\textquoteright~ this crossing ($\gamma$). 
Therefore the propagation directions are the same in one part of the paths ($L$) and
opposite in the other part (R) \cite{Sieber2001}. 

In the following we study the diagonal contribution and the off-diagonal loop corrections (Fig.\,\ref{Fig:figure8}) to $F(t)$ for graphene billiards. We find that both contributions depend on the structure of the edges due to the interplay between the boundary conditions and the (effective) TRS of graphene flakes.

\subsubsection{Diagonal contribution}

For the diagonal terms in the double sum (\ref{eq:Fsc}), the spectral form factor reads (including pairs $\gamma'=\gamma$ and $\gamma' =\gamma^{-1}$)
\begin{equation}
\label{eq:FD_sep1}
 F_D(t) = \frac{1}{2 \Th} \sum_{\gamma} |A_\gamma|^2  Z_{\gamma} \, \delta\left(T - T_\gamma\right)\,.
\end{equation}
Equation (\ref{eq:trace-K}) yields for the product of traces (\ref{eq:defZ})
\begin{eqnarray}
 Z_\gamma &=&  Z_{\gamma,\gamma}=  Z_{\gamma,\gamma^{-1}}  \\
 &=& 16
\cos^2(\theta_\gamma+ N\pi/2) \cos^2\left(2K\Lambda_\gamma + \vartheta_\gamma + N\pi/2\right) \nonumber\,,
\end{eqnarray}
provided that the total number of ac reflections $N_{\text{ac}}$ along $\gamma$ is even, otherwise $Z_\gamma=0$.
Since $\gamma$ is a periodic orbit, the total rotation angle is $2\pi w$, where $w$ is the integer winding number.
Therefore we can write
\begin{equation}
\label{eq:thetasplit}
 \theta_\gamma = (N_+ - N_-)\frac{\pi}{2} - w\pi\,,
\end{equation}
where $N_+$ and $N_-$ count the reflections with positive or negative reflection angles, respectively.
This leads to $\cos^2(\theta_\gamma + N\pi/2)=1$ 
and thus for even $N_{\text{ac}}$
\begin{equation}
\label{eq:Zgg}
Z_{\gamma} = 
16 \cos^2\left(2K\Lambda_\gamma + \vartheta_\gamma + N\pi/2 \right)\,.
\end{equation}
~
\paragraph{No ac scattering - valley conservation}~\\

For pedagogical reasons,
we first consider the simpler case of billiards whose boundaries consist only of a combination of zz and im type edges. In other words we assume that $N_{\text{ac}}=0$ for all orbits, so that there is no intervalley coupling at all.
In order to perform the orbit sum in Eq.\,(\ref{eq:Fsc}) we need to average $Z_\gamma$ accordingly. 
To perform this averaging, we first note that classically chaotic billiards exhibit ergodic dynamics, thus a long orbit will hit all boundary points with the same probability. Therefore we average $Z_\gamma$ over the possible edge types or boundary conditions, respectively. 
Since we assume that no ac edge are present $\Lambda_\gamma = 0$ in Eq.\,(\ref{eq:Zgg}). If no zz edges are present either, also $\vartheta_\gamma=0$, otherwise $\vartheta_\gamma$ is randomly distributed between zero and $2\pi$. In both cases the squared cosine in (\ref{eq:Zgg}) is on average $1/2$, since for long orbits even and odd $N$ are equally probable. That means we obtain
\begin{equation}
 \langle Z_{\gamma} \rangle = 8\,.
\end{equation}
Now we can pull $ \langle Z_{\gamma} \rangle$ out of the sum in (\ref{eq:FD_sep1}) and are left with the evaluation of the form factor as in the Schr\"odinger case. By employing the \mbox{Hannay-Ozorio de Almeida} sum rule\cite{Hannay1984} we obtain\cite{Sieber2001}
\begin{equation}
\label{eq:FD_sep2}
 F_D(t) \approx \frac{4}{\Th} D(T) |A(T)|^2  = 4t\,.
\end{equation}
According to Eq.\,(\ref{eq:FRMT2}) this result is four times
the GUE prediction for the linear term of $F(t)$. To understand the factor of four we recall the antiunitary symmetries of the problem. The physical TRS $\mathcal{T}_y$ 
would lead to the orthogonal symmetry class. However, since there is no intervalley scattering and $\mathcal{T}_y$ connects states from different valleys,
it is basically irrelevant. The effective intravalley TRS $\mathcal{T}_0$, which would lead to a symplectic ensemble, is broken by both zigzag and infinite mass
type edges. Hence we end up with two unitary subsystems that are Kramers' degenerate partners due to the remaining symmetry $\mathcal{T}_x$.
Equations (\ref{eq:Rdef}) and (\ref{eq:Fdef}) yield $F(t) = 2 F_1(2t)$ in this case, where $F_1$ is the form factor of the non-degenerate subsystem.
The loop contribution (see below) agrees with this picture.\\

\paragraph{ac billiards - complete valley mixing}~\\

We consider the opposite limit and assume that all reflections happen at ac edges. 
The term $K\Lambda_\gamma$ can only take discrete values
\begin{equation}
\label{eq:Lambda_disc}
  K\Lambda_\gamma =  \frac{\pi}{3} n \quad n \in \mathbb{Z}\,.
\end{equation}
For a generic boundary, $n$ is random and thus the average of the pseudospin contribution (\ref{eq:Zgg}) reads $\langle Z_{\gamma} \rangle = 4$. Consequently we obtain the diagonal contribution to the spectral form factor 
\begin{equation}
\label{eq:FD_mixed}
F_D(t) = 2 t\,,
\end{equation}
in agreement with the RMT prediction for the orthogonal symmetry class. The orthogonal symmetry is expected because the intervalley mixing due to the armchair scattering restores the TRS described by $\mathcal{T}_y$ \cite{Note2}
.\\

\paragraph{Mixed boundaries - the generic case}~\\

In the realistic situation of mixed boundaries, i.\,e.\,boundaries that consist of both, valley conserving edges and valley mixing edges, we obtain, based on our earlier discussion, that
\begin{equation}
 \langle Z_{\gamma} \rangle = 8\,P^{\text{e}}_{\text{ac}}(T_\gamma)\,,
\end{equation}
where $P^{\text{e}}_{\text{ac}}(T)$ denotes the average probability that an orbit with period $T$ hits armchair edges an even number of times (including zero times). Assuming chaotic dynamics and that the total length of ac edges of the billiard $W\subac = |\partial \mathcal{V}\subac|$ is small compared to the total boundary length, the ac scattering rate is approximately given by (see Ref.\,\onlinecite{Legrand1991} and references therein)
\begin{equation}
\label{eq:actime}
 \frac{1}{T\subac}  = \frac{\vF W\subac }{\pi A }\,.
\end{equation}
The \emph{armchair scattering time} $T\subac$ is the typical time between two reflections at ac type edges. In this case $P^{\text{e}}_{\text{ac}}$ is approximately given by
\begin{equation}
\label{eq:Pe_ac}
P^{\text{e}}_{\text{ac}}(T)
= \frac{1}{2}\left(1+e^{-2T/ T\subac} \right)\,.
\end{equation}
With that we can evaluate the orbit sum (\ref{eq:Fsc}) as before to obtain
\begin{equation}
\label{eq:FD_gen}
 F_D(t) = 4t P^{\text{e}}_{\text{ac}}(T) = 2 t \left(1+e^{-2 t T_{\text{H}}/T\subac} \right)\,.
\end{equation}
This formula describes the crossover between the two limiting cases of completely decoupled and mixed valleys, respectively. 
For the asymptotic limits \mbox{$T_{\text{H}}/T\subac \rightarrow 0$} and \mbox{$T_{\text{H}}/T\subac \rightarrow \infty$} we recover Eqs.\,(\ref{eq:FD_sep2}) and
 (\ref{eq:FD_mixed}), respectively. 
 
An advantage of Eq.\,(\ref{eq:FD_gen}) is that the crossover parameter does not have to be introduced phenomenologically, but can be directly related to the microscopic properties of the system, namely the total amount of (intervalley scattering) ac edges. 
The ratio of the relevant time scales is directly proportional to the total amount of ac boundary pieces, namely:
\begin{equation}
\frac{\Th}{T\subac} = 2 \kE W\subac\,.
\end{equation}
One can understand this with the following construction: if the intervalley relaxation is modeled in the spirit of B\"uttiker as intervalley relaxing lead with width $W\subac$, then ${\Th}/{T\subac}$ is proportional to the number channels of the intervalley relaxing lead.

\begin{figure}
 \centering
\includegraphics[width=0.48\textwidth]{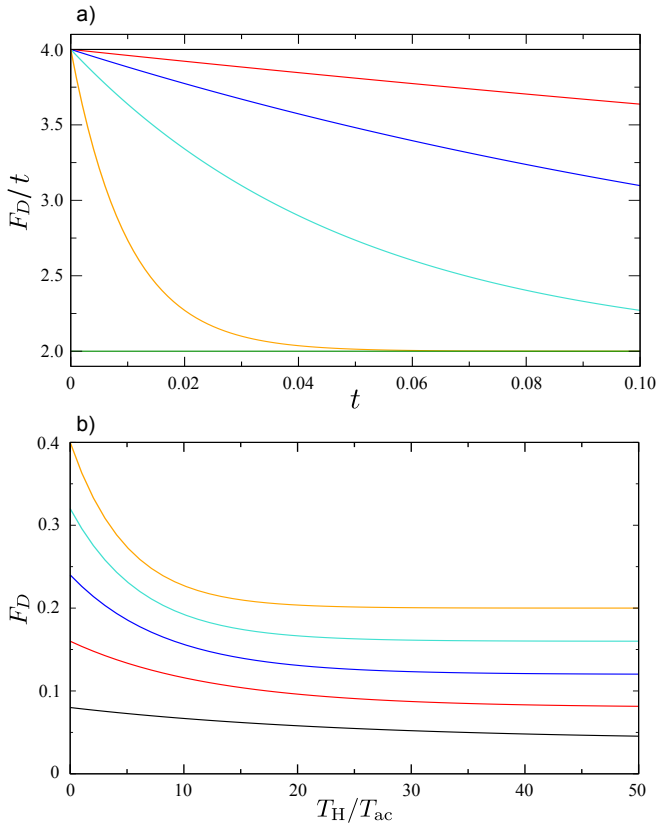}
\caption{\small{a) Rescaled diagonal contribution $F_D(t)/t$ to the spectral form factor as a function of $t=T/\Th$ for a generic graphene cavity for different values of the intervalley scattering strength. From top to bottom:
$\Th/T\subac = 0$\,(black), $1$\,(red), $3$\,(blue), $10$\,(turquoise), $50$\,(orange) and $\infty$\,(green).
While in the limiting cases of $\Th/T\subac \rightarrow 0$ and $\Th/T\subac \rightarrow \infty$
the $t$-dependence of $F_D$ is purely linear, this is not generally the case. For all finite values of $\Th/T\subac$, $F_D$ shows signatures of 
unitary correlations at small $t$.
b) From bottom to top, $F_D$ as a function of $\Th/T\subac$ at $t=0.02$\,(black), $0.04$\,(red), $0.06$\,(blue), $0.08$\,(turquoise) and $0.1$\,(orange). 
It is an exponential crossover from 4 times the GUE to the GOE result.}}
\label{Fig:FD}
\end{figure}

Figure \ref{Fig:FD}\,a) shows $F_D(t)/t$ for small $t=T/\Th$ and the parametric crossover between the limiting RMT symmetry classes, namely GOE [lowest line, Eq.\,(\ref{eq:FD_mixed})] and $4\times$GUE [uppermost line, Eq.\,(\ref{eq:FD_sep2})]. In the general case of finite $\Th/T\subac$, 
$F_D$ is not purely linear as in both limiting cases. Note that even for a considerable amount of ac scattering (consider e.\,g.\,the (turquoise) curve for $\Th/T\subac=10$) we still find rather strong deviations from the GOE statistics. Panel b) shows the crossover of $F(t)$ as a function of $\Th/T\subac$.
We point out that this crossover from 4$\times$GUE to GOE is rather particular: Usually transitions between GUE and 
GOE result from a symmetry \emph{breaking} mechanism, e.\,g.\,due to a magnetic field. Here the armchair edges and the resulting intervalley scattering act 
as a symmetry \emph{restoring} mechanism. Moreover the correlations in the unitary limit are \emph{weaker} than in the orthogonal case, in contrast to the case of 
the usual GOE-GUE transition.

In a mesoscopic graphene quantum dot the crossover parameter $\Th/T\subac$ can be tuned by changing the energy. Since our theory is valid in a wide range of $\kE$, the transition in Fig.\,\ref{Fig:FD}\,b) should be accessible in a real system. Consider for example a structure with a typical length scale of $1000\,a$, then we expect our theory to be valid for $0.01 \lesssim \kE a \lesssim 0.2$, approximately. Assuming $W\subac \approx 250\,a$ this leads to $5 \lesssim \Th/T\subac \lesssim 100$.

\subsubsection{Off-diagonal (loop) contributions}
\label{ssec:F_loops}
Now we proceed with the calculation of the correction terms (order $t^2$) to the form factor. At this order it is the loop pairs $(\gamma,\gamma_\times)$ depicted in Fig.\,\ref{Fig:figure8}, that contribute to $F(t)$. We denote these contributions by $F_L$. Along the two loops $L$ and $R$ the two trajectories are exponentially close to each other. For vanishing Ehrenfest time, we assume that the sequences of reflections along $\gamma$ and $\gamma_\times$ are exactly equal in the $L$-part while they are equal but with opposite order in the $R$-part. Orbit pairs that differ in their number of reflections are of measure zero in the semiclassical limit. We start by calculating the pseudospin propagators. All quantities will be labeled correspondingly for the loops $L,R$. 
For the total reflection angles we have
\begin{equation}
\label{eq:sep_theta}
 \theta_\gamma = \theta_L + \theta_R\,, \quad \theta_{\gamma_\times} = \theta_L - \theta_R\,.
\end{equation}
As we show in App.\,\ref{app:pspintrace_2}, the expression (\ref{eq:trace-K}) for the traces leads to 
\begin{eqnarray}
 \label{eq:TrKgTrKgx}
 Z_{\gamma,\gamma_\times} &=&
 (-1)^{N_R+1}16 \cos\left(2K\Lambda_\gamma + \vartheta_\gamma +N \pi/2\right)  \nonumber \\ &\times& 
\cos\left(2K\Lambda_{\gamma_\times} + \vartheta_{\gamma_\times}+N \pi/2\right) 
\end{eqnarray}
if $N\subac$ is even for both orbits, and $Z_{\gamma,\gamma_\times}=0$ otherwise.\\

\paragraph{No ac scattering - valley conservation}~\\

When there is no ac scattering, the valleys are uncoupled and we have
\begin{equation}
 \label{eq:ZL_zz}
Z_{\gamma,\gamma_\times}
 =  (-1)^{N_R+1}16\cos\left(\vartheta_\gamma+N \pi/2 \right)
 \cos\left(\vartheta_{\gamma_\times}+N \pi/2\right)\,.
\end{equation}
We decompose the angles $\vartheta$ into $\vartheta_L \pm \vartheta_R$ and thus obtain
\begin{eqnarray}
\label{eq:trig_1}
&& \cos\left(\vartheta_\gamma +N \pi/2\right)
 \cos\left(\vartheta_{\gamma_\times}+N \pi/2\right) \\  &=&
  \cos^2\left(\vartheta_L+N \pi/2\right)\cos^2\left(\vartheta_R\right) \nonumber 
\\ &-& \sin^2\left(\vartheta_L+N \pi/2\right)\sin^2\left(\vartheta_R\right) \nonumber \,.
\end{eqnarray}
The averaged pseudospin contribution to $F_L$ vanishes
\begin{equation}
\label{eq:ZavL_1}
 \langle Z_{\gamma,\gamma_\times} \rangle = 0\,,
\end{equation}
because the angles  $\vartheta_L$ and $\vartheta_R$ have to be averaged separately, and thus the cosine terms and the sine terms cancel on average. This result holds also if zz edges are absent and all reflections happen at im edges. In the latter case $\vartheta_\gamma =  \vartheta_{\gamma_\times} = 0$ and therefore $Z_{\gamma,\gamma_\times}$ is either identically zero (odd $N$) or $Z_{\gamma,\gamma_\times}\sim (-1)^{N_R}$ (even $N$), leading again to Eq.\,(\ref{eq:ZavL_1}), because for long orbits there are equally many orbits with even and odd $N_R$\cite{Berry1987}. That means there is \emph{no loop contribution} to the spectral form factor,
\begin{equation}
\label{eq:FL_sep}
F_L(t) =0 \,,
\end{equation}
if there are no intervalley scattering ac edges.\\

\paragraph{ac billiards - complete valley mixing}~\\

In the opposite limit of dominant ac scattering, Eq.\,(\ref{eq:TrKgTrKgx}) gives for even $N=N\subac$
\begin{eqnarray}
Z_{\gamma,\gamma_\times} =  (-1)^{N_R+1}16\cos\left(2K\Lambda_\gamma\right)
 \cos\left(2K\Lambda_{\gamma_\times}\right) \,.
\end{eqnarray}
To average this, we have to distinguish between two different cases.
First we assume that $N_R$, the number of reflections in the $R$-part of $\gamma$, is even. Then $\Lambda_L$ and $\Lambda_R$ are well defined and we have 
\begin{equation}
\Lambda_{\gamma} = \Lambda_L + \Lambda_R\,, \quad \Lambda_{\gamma_\times} = \Lambda_L - \Lambda_R\,.
\end{equation}
From the trigonometric relation analog to Eq.\,(\ref{eq:trig_1}) it follows that these orbits do not contribute on average. 
On the other hand if $N_R$ is odd, 
it is straight-forward to show that 
\begin{equation}
 \cos(2K\Lambda_{\gamma_\times}) = \cos(2K\Lambda_{\gamma})
\end{equation}
and therefore for even $N$ 
\begin{equation}
 Z_{\gamma,\gamma_\times} = 16 \cos^2(2K\Lambda_{\gamma}) \,.
\end{equation}
Since only orbits with even $N$ and odd $N_R$ contribute we obtain for the average
\begin{equation}
\langle Z_{\gamma,\gamma_\times}\rangle = 4\langle \cos^2(2K\Lambda_{\gamma}) \rangle = 2\,,
\end{equation}
that is the graphene specific pseudospin trace enters merely with an overall prefactor 2 to the spectral form factor. The rest of the calculation is identical to evaluation of the semiclassical form factor in Ref.\,\onlinecite{Sieber2001}. The full contribution of pairs of orbits $(\gamma, \gamma_\times)$ is obtained by summing over all $\gamma$ and for each $\gamma$ counting the number of self encounters, i.\,e.~self intersections with a (small) crossing angle $\varepsilon$. We introduce $P(\varepsilon, T_\gamma)$ as a measure for the density of self crossings of an orbit of length $T_\gamma$\cite{Sieber2001}, 
\begin{equation}
\label{eq:LoopDens}
 P(\varepsilon,T) \approx  \frac{\vF^2\,\varepsilon}{\pi A} \intl_{\Tmin}^{T-\Tmin}\!\! dT_R\, [T-T_R]\,.
\end{equation}
Here
\begin{equation}
\label{eq:Tmin}
 \Tmin = -\frac{2}{\lambda} \ln(c\, \varepsilon)
\end{equation}
with $\lambda$ the average Lyapunov exponent and $c$ a constant, is the minimal time to form a closed loop.
Then the angular integral over the exponents of the phase differences \mbox{$\delta S_{\gamma,\gamma_\times}/\hbar$} is computed, leading to\cite{Sieber2001}
\begin{equation}
\label{eq:FL_ac}
\mathrm{Re}\intl_0^\pi d\varepsilon \, e^{i\,\delta S_{\gamma,\gamma_\times}/\hbar} P(\varepsilon, T_\gamma) = -t 
\end{equation}
as for a Schr\"odinger billiard\cite{Sieber2001}. Evaluating the remaining orbit sum similar to Eq.\,(\ref{eq:FD_sep1}) yields
\begin{eqnarray}
\label{eq:FL_mixed}
  F_L(t) = -2\, t^2\,,
\end{eqnarray}
in agreement with the RMT prediction [cf. Eq.\,(\ref{eq:FRMT2})] for the GOE.\\

\paragraph{Mixed boundaries - the generic case}~\\

We now consider the general case, where all three types of edges - ac, zz, and im - are present.  From the previous discussion we know that the contribution to $F_L$ is exclusively due to such classical orbits $\gamma$ that undergo odd numbers of ac reflections along both, the $L$ ($N_{\text{ac},L}$) and the $R$ ($N_{\text{ac},R}$) parts of $\gamma$. For these orbits we obtain $\langle Z_{\gamma,\gamma_\times}\rangle = 8$ also in the presence of zz and im type reflections, while $\langle Z_{\gamma,\gamma_\times}\rangle =0$ for the remaining orbits. Therefore we can write, analogous to Eq.\,(\ref{eq:FL_ac}),
\begin{equation}
\label{eq:FL_all}
  F_L(t) = 8\, t\, \mathrm{Re}\intl_0^\pi d\varepsilon \, e^{i\,\delta S_{\gamma,\gamma_\times}/\hbar} \tilde{P}(\varepsilon, T_\gamma)
\end{equation}
with $\tilde{P}(\varepsilon, T_\gamma)$ the corresponding density of self crossings.
In order to determine the density of self crossings of orbits that fulfill the condition of odd $N_{\text{ac},L/R}$, we start from the unrestricted density of loop pairs (\ref{eq:LoopDens}).
To account for the fact that only the subset of loop pairs with odd $N_{\text{ac},L/R}$ contributes, we have to modify $P$ to obtain
\begin{eqnarray}
\label{eq:modLoopdens}
 \tilde{P}(\varepsilon,T) &\approx& \frac{\vF^2\,\varepsilon}{\pi A} \intl_{\Tmin}^{T-\Tmin}\!\! dT_R\, [T-T_R] \\
&&\times P\supt{o}\subac[T_R-\Tmin] P\supt{o}\subac[T-T_R-\Tmin] \nonumber
\end{eqnarray}
with the approximate probability to hit ac edges an odd number of times during a time $T$
\begin{equation}
\label{eq:Pac_o}
 P\supt{o}\subac(T) = \frac{1}{2}\left(1-e^{-2\, T /T\subac}\right)\,.
\end{equation}
Using Eq.\,(\ref{eq:Pac_o}) when evaluating the integral in Eq.\,(\ref{eq:modLoopdens}) yields
\begin{eqnarray}
\tilde{P} (\varepsilon,T_\gamma) &\approx& \frac{\vF^2\,\varepsilon}{\pi A} \frac{T}{8}
 \bigg( T-T\subac  + (T+T\subac)e^{-2T/T\subac}  \\  
&&  -2\Tmin \left[ 1-\left(1+\frac{2T}{T\subac}\right)e^{-2T/T\subac}  \right] \bigg)\,.  \nonumber
\end{eqnarray}
Here we assumed that (in the RMT limit) $\Tmin$ is much shorter than the other time scales $T$ and $T\subac$ and therefore neglected terms of higher order in $\Tmin$. The $\varepsilon$-independent terms of $\tilde{P}$ do not contribute\cite{Sieber2001} to the real part of the integral in Eq.\,(\ref{eq:FL_all}), so that we get
\begin{equation}
\label{eq:FL_gen}
 F_L(t) \approx -2t^2  \left[ 1-\left(1+\frac{2t\Th}{T\subac}\right)e^{-2t\Th/T\subac}  \right]\,.
\end{equation}
We recover the asymptotic limits (\ref{eq:FL_sep}) and (\ref{eq:FL_mixed}) for \mbox{$T_{\text{H}}/T\subac \rightarrow 0$} and \mbox{$T_{\text{H}}/T\subac \rightarrow \infty$}, respectively.

\begin{figure}
 \centering
\includegraphics[width=0.48\textwidth]{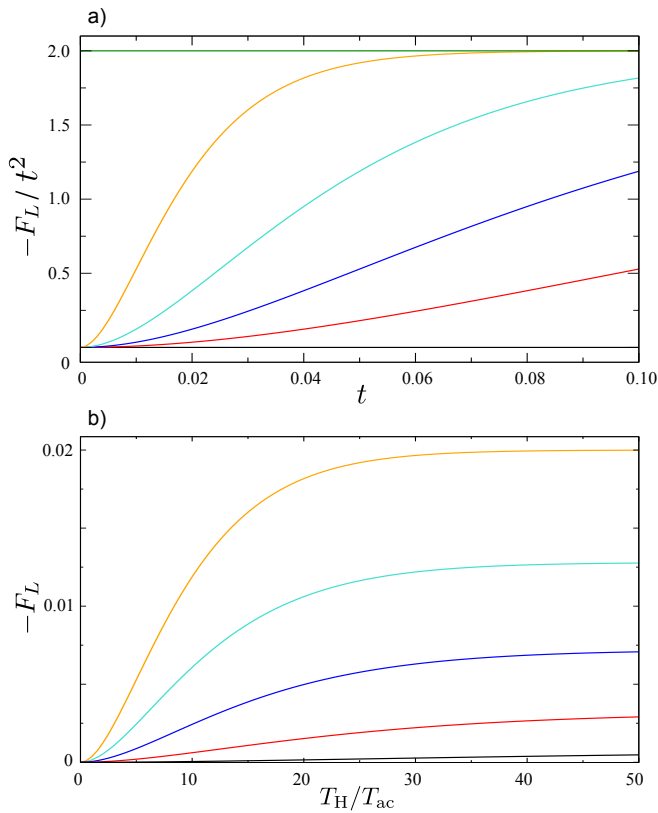}
\caption{\small{a) Absolute value of the rescaled leading off-diagonal contribution, $-F_L(t)/t^2$, to the form factor as a function of $t=T/\Th$ for different values (from bottom to top) of the ac scattering strength
$\Th/T\subac = 0$\,(black), $5$\,(red), $10$\,(blue), $20$\,(turquoise), $50$\,(orange) and $\infty$\,(green). 
At $\Th/T\subac=0$ $F_L$ is vanishing, in agreement with GUE, while for larger values the curve approaches the quadratic behavior of the GOE prediction.
b) Exponential crossover of $-F_L$ as a function of $\Th/T\subac$ (from bottom to top)
at $t=0.02$\,(black), $0.04$\,(red), $0.06$\,(blue), $0.08$\,(turquoise) and $0.1$\,(orange). }}
\label{Fig:FL}
\end{figure}

Figure~\ref{Fig:FL}\,a) shows $-F_L(t)/t^2$ for various effective ac scattering strengths $\Th/T\subac$ in the range of small $t$, where $F_L$ is the dominant off-diagonal contribution to the form factor. The parameter $\Th/T\subac$ controls the crossover between the two RMT limits: $F_L=0$ for $\Th/T\subac \rightarrow 0$ and $F_L=-2t^2$ for $\Th/T\subac \rightarrow \infty$. In panel b) we plot the crossover of $-F_L(t)$ as a function of $\Th/T\subac$.

Equations (\ref{eq:FD_gen}) and (\ref{eq:FL_gen}) are the main results of this section. They describe the spectral correlations of chaotic graphene billiards with a boundary that consists of an arbitrary combination of ac, zz and im type edges. The total amount of intervalley scattering at armchair edges sets the time scale $T\subac$ that represents the control parameter for a $4\times$GUE to GOE crossover. In App.\,\ref{app:FwM} we generalize our results by incorporating direct TRS breaking, e.\,g.\,by including  a small magnetic flux $\Phi$ through the billiard. Due to the flux induced Aharonov-Bohm type phase differences, this causes a breaking of both, the TRS $\mathcal{T}_y$ and the valley symmetry $\mathcal{T}_x$ leading to another crossover. In the asymptotic cases of vanishing ac edges and complete valley mixing, we find transitions $4\times$GUE$\rightarrow 2\times$GUE and GOE$\rightarrow$ GUE, respectively.

%

\section{Transport through open graphene cavities}
\label{sec:transport}

The crossover between the different effective universal symmetry classes discussed for spectral statistics in the preceeding chapter is also reflected in quantum transport properties that are experimentally more directly accessible.
In this part of the paper we focus on the two terminal conductance of open graphene cavities. Our starting point is the linear response expression for the conductance from lead $b$ to lead $a$ in terms of the Green's function of the cavity\cite{Baranger1989}
\begin{equation}
\label{eq:conductance}
 g_{ab} = -\int\limits_{C_a}dy\int\limits_{C_b}dy'~ \sigma_{ab}(\bsy{x},\bsy{x'})\,,
\end{equation}
where $C_a$ and $C_b$ are the cross-sections of the leads and $\sigma_{ab}$ is the non-local conductivity for graphene\cite{Baranger1989, Nomura2007, Crepieux2001},
\begin{equation}
\label{eq:conductivity}
\sigma_{ab}(\bsy{x},\bsy{x'}) = \frac{e^2}{2\pi\hbar} \tr\left[\sigmaa G(\bsy{x},\bsy{x'}) \sigmab G^{\dagger}(\bsy{x},\bsy{x'})\right]\,.
\end{equation}
Here, the vectors $\bsy{a}$ and $\bsy{b}$ are unit vectors in the direction of the corresponding lead and point into the interior of the system (cf. Fig.\,\ref{Fig:loops}). 
In App.\,\ref{app:FisherLee} we show that this method to compute $g_{ab}$ is equivalent to an approach based on a Fisher-Lee type formula\cite{Fisher1981} generalized to graphene, which we also derive in App.\,\ref{app:FisherLee}.

Inserting the semiclassical Green's function (\ref{eq:Gsc})
into Eq.\,(\ref{eq:conductivity}), we obtain a double sum over classical orbits $\gamma$ and $\gamma'$ that lead from point $\bsy{x'}$ in lead $b$
to point $\bsy{x}$ in lead $a$:
\begin{equation}
\label{eq:sigma_sc}
\sigma_{ab}(\bsy{x},\bsy{x'}) \approx \left(\frac{ e \vF}{4\pi\hbar}\right)^2  
\sum_{\gamma \gamma'} 
Y_{\gamma,\gamma'}
D_{\gamma}D_{\gamma'} 
e^{i \delta S_{\gamma,\gamma'}/\hbar+i\frac{\pi}{2}\delta\mu_{\gamma,\gamma'}}
\end{equation}
with $\delta S_{\gamma,\gamma'} = \hbar\kE(L_\gamma - L_{\gamma'})$, $\delta \mu_{\gamma,\gamma'} = \mu_\gamma - \mu_{\gamma'}$ and
\begin{equation}
  Y_{\gamma,\gamma'} = \tr\left(\sigmaa \Ktilde_{\gamma} \sigmab \Ktilde^\dagger_{\gamma'}\right)\,. 
\end{equation}
We note that the entire effect of the graphene pseudospin and valley dynamics is included in the factor $Y_{\gamma,\gamma'}$. The main difference of this factor from $Z_{\gamma,\gamma'}$ in Eq.\,(\ref{eq:Fsc}) is that the pseudospin propagators have to be multiplied \emph{before} tracing.

\subsection{The average quantum conductance}
\subsubsection{Quantum chaotic Schr\"odinger billiards}

First we focus on the average conductance, \mbox{$\langle g_{ab} \rangle_{\kE} = \langle g_{ab} \rangle$} where as in Sec.\,\ref{sec:spec-stat}  the averaging is performed over an energy window that is classically small but still contains many quantum levels. Based on semiclassical approaches for the Schr\"odinger case\cite{Baranger1993, Richter2002, Heusler2006, Zaitsev2005, Bolte2007} and random matrix theory\cite{Baranger1994, Jalabert1994} coinciding universal predictions for the average quantum conductance were made for chaotic cavities for the unitary, orthogonal, and symplectic universality class:
\begin{equation}
 \frac{\langle g_{ab}\rangle}{g_0}  = \frac{M_a M_b}{M_a + M_b - 1 + 2/\beta } \,.
\end{equation}
Here $\beta$ depends on the universality class and the considered random matrix ensemble, respectively, namely $\beta = 1$ for the circular orthogonal ensemble (COE),  $\beta=2$ for the circular unitary ensemble (CUE) and $\beta=4$ for the circular symplectic ensemble (CSE).
The conductance quantum is defined as $g_0 = e^2/h$
and $M_{a/b}$ is the number of propagating channels in the corresponding lead.
For large $M_{a/b}$, we can expand $\langle g_{ab}\rangle$ as
\begin{equation}
\label{eq:gRMT_expansion}
 \frac{\langle g_{ab}\rangle}{g_0} = \frac{M_a M_b}{M_a + M_b} + \left(1- \frac{2}{\beta} \right) 
\frac{M_a M_b}{(M_a + M_b)^2} + \mathcal{O}\left(M_{a/b}^{-1}\right) \,.
\end{equation}
Semiclassically, just as for the spectral form factor, relevant contributions to the average conductance (\ref{eq:gRMT_expansion}) are due to orbit pairs that have a small or vanishing action difference $\delta S_{\gamma,\gamma'}$. In fact for generic pairs with $\delta S_{\gamma,\gamma'} \sim \hbar \kE L$ the fastly oscillating exponential in (\ref{eq:sigma_sc}) leads to zero average conductance. Here, similar to the case of $F(t)$, we compute contributions to $\langle g_{ab}\rangle$ due to diagonal and leading off-diagonal terms in Eq.\,(\ref{eq:sigma_sc}) in the next two subsections. 

\subsubsection{Diagonal contribution}

We begin with the diagonal contribution, where $\gamma'=\gamma$ and $\delta S_{\gamma,\gamma'}= \delta \mu_{\gamma,\gamma'}=0$. We note that the boundary matrices in Eq.\,(\ref{eq:K_i}) do not mix valley and sublattice pseudospin. Thus we can separate the graphene pseudospin propagator into a sublattice part $\Ktildes_{\gamma}$ and a unitary valley part $\Ktildev_{\gamma}$, which we trace out immediately giving $\tr\left(\Ktildev_{\gamma}  \Ktildevd_{\gamma}\right)=2$. The pseudospin contribution in diagonal approximation then becomes
\begin{eqnarray}
 Y_{\gamma} &=&
\tr\left(\sigmaa \Ktildes_{\gamma}\sigmab  \Ktildesd_{\gamma}\right)  
 \tr\left(\Ktildev_{\gamma}  \Ktildevd_{\gamma}\right) \nonumber \\
&=& 2\, \tr\left(\sigmaa \Ktildes_{\gamma}\sigmab  \Ktildesd_{\gamma}\right)\,.
\end{eqnarray}
The sublattice part depends on the angles of the incoming and outgoing trajectory segments (cf. Fig.\ref{Fig:loops}) leading to
\begin{equation}
\label{eq:Ydiag}
Y_{\gamma} = - 8 \cos(\theta_a)\cos(\theta_b)\,,
\end{equation}
as shown in App.\,\ref{app:pspintrace_3}.
This result for the pseudospin trace within diagonal approximation holds irrespective of the specific edge types involved in the various reflections along $\gamma$ (in contrast to the diagonal contributions to the spectral form factor). 
We now compute the full diagonal contribution to Eq.\,(\ref{eq:sigma_sc}) as single sum over orbits with fixed outgoing and incoming angles employing the classical sum rule \cite{Sieber1999} 
\begin{equation}
\label{eq:sumrule}
 \sum_{\gamma (\theta_a,\theta_b)}\!\! |D_\gamma|^2 \,\delta(T-T_\gamma) = \frac{\hbar \kE^2}{\vF \Sigma(\kE)} d\theta_ad \theta_b e^{-T/\Td} 
\end{equation}
to transform the sum into a triple integral over the time the particle spends in the cavity and entrance and exit angles. Here $\Td$ is the dwell time, i.\,e.\,the time a classical particle typically spends within the cavity. In terms of the cavity area and the lead widths $W_a$ and $W_b$, the corresponding escape rate is approximately given by (see Ref.\,\onlinecite{Legrand1991} and references therein)
\begin{equation}
\label{eq:dwelltime}
\frac{1}{\Td} \approx \frac{\vF}{\pi A}(W_a+W_b)\,.
\end{equation}
Additionally we find for the energy surface $\Sigma$ in Eq.\,(\ref{eq:sumrule})
\begin{equation}
\label{eq:Esurf}
 \Sigma(\kE) = \int d^2x \int d^2k \,\delta(\kE - k) = 2\pi A \kE\,.
\end{equation}
Combining Eqs.\,(\ref{eq:sigma_sc}, \ref{eq:Ydiag}-\ref{eq:Esurf}) we get for the diagonal non-local conductivity
\begin{eqnarray}
 \langle \sigma_D (\bsy{x},\bsy{x'})\rangle &=& -\left(\frac{e\vF}{4\pi\hbar}\right)^2\intl_{-\pi/2}^{\pi/2} \!\! d\theta_a \!\! \intl_{-\pi/2}^{\pi/2} \!\!d\theta_b \,
  8 \cos(\theta_a)\cos(\theta_b) \nonumber \\
 && \times \frac{\hbar \kE}{\vF  2\pi A} \int_0^\infty dT\, e^{-T/\Td} \\ \nonumber &=& -\frac{2\kE}{\pi}\frac{g_0}{W_a+W_b}\,. \\
\end{eqnarray}
After integration over the lead openings in Eq.\,(\ref{eq:conductance}) we obtain
\begin{equation}
\label{eq:gD}
 \frac{\langle g_D \rangle}{g_0} = \frac{2\kE}{\pi}\frac{W_a W_b}{W_a+W_b} \approx  \frac{M_a M_b}{M_a+M_b} \,,
\end{equation}
without edge signatures and in agreement with the leading order term in expression (\ref{eq:gRMT_expansion}). In Eq.\,(\ref{eq:gD}) we have approximated the numbers of propagating lead channels $M_a$ and $M_b$ by
\begin{equation}
 M_{a/b} \approx \frac{2\kE}{\pi} W_{a/b}\,,
\end{equation}
accounting for the two valleys.
For identical leads ($M_a=M_b=M$) we have $\langle g_D  \rangle /g_0 =M/2$, half of the maximum possible conductance, as expected 
for a classical particle being randomly scattered.

\subsubsection{Loop contributions: Weak localization}
\begin{figure}
 \centering
 \includegraphics[width=0.35\textwidth]{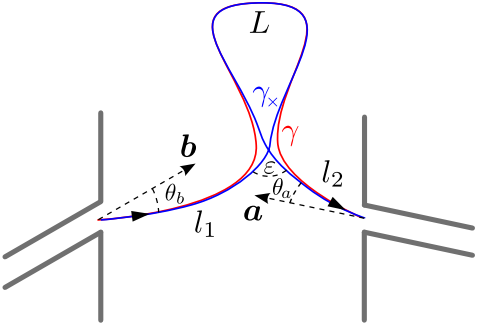}
\caption{\small{Sketch of a loop pair that contributes to the weak localization correction to the average conductance\cite{Richter2002}. One of the orbits 
crosses itself with an angle $\varepsilon$, the other orbit avoids this crossing but apart from this follows the first one exponentially close.}}
\label{Fig:loops}
\end{figure}
\begin{figure}
 \centering
 \includegraphics[width=0.48\textwidth]{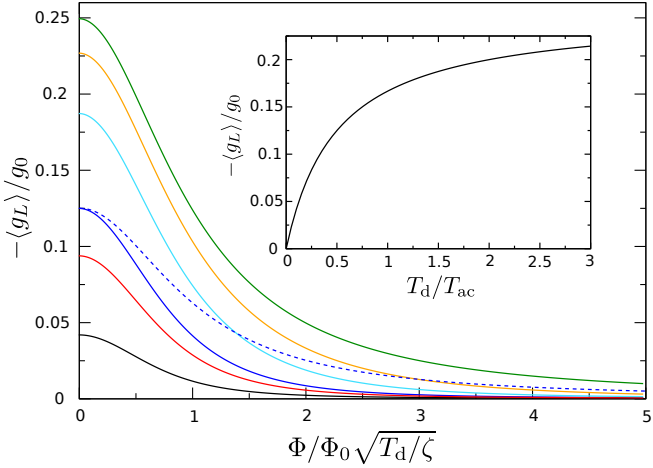}
\caption{\small{Absolute value of the WL correction, $-\langle g_L \rangle$, as a function of the magnetic flux $\Phi$ through the system, obtained 
from our semiclassical theory for identical leads.
The full lines correspond to different values of the effective armchair scattering strength $\Td/T\subac$ (from bottom to top) : 0.1 (black), 0.3 (red), 0.5 (blue), 1.5 (turquoise), 
5 (orange) and $\infty$ (green). The dashed blue line shows a Lorentzian curve $\sim \left[1+ (\Phi^2 \Td)/(\Phi_0^2 \zeta)\right]^{-1}$for comparison.
The inset shows the crossover as a function of $\Td/T\subac$ at zero magnetic flux according to Eq.\,(\ref{eq:gL_zeroPhi}).
}}
\label{Fig:gL}
\end{figure}

We now evaluate the leading order quantum correction to the average conductance, namely the loop contributions that are responsible for the \textit{weak localization} (WL) correction in $\langle g_{ab} \rangle$. The corresponding orbit pairs $(\gamma,\gamma_\times)$, sketched in Fig.\,\ref{Fig:loops}, are similar to the ones considered in Sec.\,\ref{sec:spec-stat}; however here the orbits are not periodic but they are open orbits connecting one lead to the other. We divide the orbits into the legs $l_1$ and $l_2$ where $\gamma$ and $\gamma_\times$ both have the same propagation direction and 
the loop part $L$, where the propagation directions are opposite (cf. Fig.\,\ref{Fig:loops}), and label all quantities correspondingly. For vanishing Ehrenfest time, the effects of a finite encounter time can be neglected here.
In App.\,\ref{app:pspintrace_4}, we compute the trace of the pseudospin matrices of the loop pairs and find
\begin{eqnarray}
\label{eq:Yggx}
Y_{\gamma,\gamma_\times} &=& 
8\cos(\theta_a)\cos(\theta_b)  \\
&\times& \left\{
\begin{array}{cl}
 -1 & \text{for~odd~}N_{\text{ac},L}\,, \\
 (-1)^{N_L} \cos(4K \Lambda_L + 2\vartheta_L) & \text{for~even~}N_{\text{ac},L}\,.
\end{array}
\right. \nonumber
\end{eqnarray}
For the summation in Eq.\,(\ref{eq:sigma_sc}), $Y_{\gamma,\gamma_\times}$ has to be averaged according to the edge characteristics of the system.
We make use of the ergodic dynamics of classically chaotic cavities, particularly the fact that long orbits hit all points on the boundary with the same probability. Therefore we average $Y_{\gamma,\gamma_\times}$ over the boundary conditions corresponding to the different edge types.
Here again the ac edges play a crucial role: Only classical orbits that hit ac edges an odd number of times during the loop part lead to a finite contribution to the average conductance. For pairs with odd $N_{\text{ac},L}$
we have an average pseudospin contribution \mbox{$\langle Y_{\gamma,\gamma_\times}\rangle = -8\cos(\theta_a)\cos(\theta_b)$}, as follows directly from Eq.\,(\ref{eq:Yggx}). On the other hand for even $N_{\text{ac},L}$ we obtain \mbox{$\langle Y_{\gamma,\gamma_\times}\rangle =0$}\cite{Note3}. To see this we first assume that all reflections occur at ac edges, i.\,e. $N_L=N_{\text{ac},L}$ and $\vartheta_L=0$. Since for a generic boundary $K\Lambda_L = n\pi/3$ [cf. Eq.\,(\ref{eq:sigma_sc})] with $n$ a random integer, the cosine in Eq.\,(\ref{eq:Yggx}) is zero on average. The result of this averaging is not altered by the inclusion of reflections from zz and im edges.
Using the sum rule (\ref{eq:sumrule}) and replacing the $\gamma$ sum by integrations over $T$, $\theta_a$, $\theta_b$, and the crossing angles $\varepsilon$, we find then for the loop correction to the non-local conductivity 
\begin{eqnarray}
\label{eq:sigL_ints}
 \langle \sigma_L (\bsy{x},\bsy{x'}) \rangle &=& -\frac{2e^2\vF\kE}{\pi^3\hbar A}
 \mathrm{Re}\intl_{0}^\pi d\varepsilon\!\!\!\! \intl_{2\Tmin}^\infty \!\!\!\!dT\, e^{-[T-\Tmin]/\Td} \nonumber \\ 
 && \times\check{P}(\varepsilon,T) e^{i\delta S_{\gamma,\gamma_\times}/\hbar} \,. 
\end{eqnarray}
Note that taking the real part and including an additional factor of two accounts for the fact that every orbit pair in the sum has a partner pair, where $\gamma$ and $\gamma_\times$ are interchanged leading to the complex conjugated term.
In Eq.\,(\ref{eq:sigL_ints}) $\check{P}$ denotes the density of loop pairs that hit ac edges an odd number of times along the loop $L$. A weak magnetic field that causes a flux $\Phi$ through the area of the cavity leads to dephasing on a timescale
\begin{equation}
\label{eq:magtime}
 \TB = \zeta \frac{\Phi_0^2}{\Phi^2}\,,
\end{equation}
with the magnetic flux quantum  $\Phi_0 = h / e$ and a system specific parameter $\zeta$, see Eq.\,(\ref{eq:TB}). In App.\,\ref{app:FwM} we show how this affects $\check{P}$. The modified loop density is then given by
\begin{eqnarray}
\label{eq:modLoopdens_g}
\check{P}(\varepsilon,T) \approx && \hspace*{-0.2cm} \frac{\vF^2 \varepsilon}{2\pi A} \intl_{\Tmin}^{T-\Tmin}\!\!\!\! dT_L\, [T-T_L-\Tmin]  \\
&\times& e^{-[T_L-\Tmin]/\TB} \left(1- e^{-[T_L-\Tmin]/T\subac}\right)\,,\nonumber 
\end{eqnarray}
where $T\subac$ is the typical time between two reflections from intervalley scattering ac edges [see Eq.\,(\ref{eq:actime})] and we used the probability for an odd number of ac reflections (\ref{eq:Pac_o}). Further $T\subt{min}$ is the minimum time to form a closed loop [see Eq.\,(\ref{eq:Tmin})].
Evaluating the integrals in Eq.\,(\ref{eq:sigL_ints}) in the limit $\lambda \Td \gg \ln(\vF \kE /\lambda)$, we find\cite{Richter2002, Waltner2010} 
\begin{eqnarray}
\label{eq:gL}
\hspace*{-0.7cm}
\frac{ \langle g_L \rangle}{g_0} 
&\approx&   - \frac{M_a M_b}{(M_a+M_b)^2} \left(\frac{1}{1+\frac{\Td}{\TB}} -
 \frac{1}{1+\frac{\Td}{\TB}+2\frac{\Td}{T\subac}}\right) . 
\end{eqnarray}
This central result represents the leading order quantum correction in inverse channel number (known as weak localization correction) to the average magnetoconductance of a ballistic graphene cavity. Three competing timescales govern $\langle g_L \rangle$: the armchair scattering time $T\subac$, Eq.\,(\ref{eq:actime}), the dwell time $\Td$, Eq.\,(\ref{eq:dwelltime}), and the magnetic time $\TB$, Eq.\,(\ref{eq:magtime}).
Equation (\ref{eq:gL}) describes the WL - no WL crossover as a function 
of the amount of armchair scattering and the magnetic field. Unlike for the spectral form factor, Eq.\,(\ref{eq:FL_gen}), there is no dependence on the Fermi momentum $\kE$ (for fixed numbers of propagating lead channels).
For vanishing $\Phi$ the size of the WL is given by
\begin{equation}
\label{eq:gL_zeroPhi}
 \lim_{\Phi\rightarrow 0} \frac{\langle g_L \rangle}{g_0} \approx - \frac{M_a M_b}{(M_a+M_b)^2}  \frac{1}{1+T\subac/(2\Td)}\,,
\end{equation}
describing the $T\subac$ dependence of $\langle g_L \rangle$, cf. inset in Fig.\,\ref{Fig:gL}.\\

\paragraph{ac billiards - complete valley mixing}~\\ 

For very strong armchair scattering, $T\subac \ll \Td$, the well known Lorentzian magnetoconductance profile\cite{Baranger1993, Richter2002} is restored,
\begin{equation}
\lim_{T\subac \rightarrow 0} \frac{\langle g_L \rangle}{g_0} \approx - \frac{M_a M_b}{(M_a+M_b)^2}  \frac{1}{1+\Td/\TB}\,,
\end{equation}
describing the usual GOE $\rightarrow$ GUE transition, where the total WL signal has a size of approximately $-1/4$ for identical leads in the limit of large channel numbers. Since the valleys are completely mixed, the TRS $\mathcal{T}_y$ is active at $\Phi=0$, leading to the orthogonal universality class. The magnetic field breaks $\mathcal{T}_y$, driving the system into the unitary class.\\

\paragraph{Weak ac scattering}~\\

For very weak ac scattering,  $T\subac \gg \Td$, Eq.\,(\ref{eq:gL}) becomes
\begin{equation}
\label{eq:gL_weak_ac}
\lim_{T\subac \rightarrow \infty} \frac{\langle g_L \rangle}{g_0} \approx - \frac{M_a M_b}{(M_a+M_b)^2}  \frac{1}{(1+\Td/\TB)^2} \frac{2\Td}{T\subac}\,,
\end{equation}
that is the WL correction vanishes linearly with $\Td/T\subac$, reflecting (partial) effective TRS breaking. The overall TRS associated with the antiunitary operator $\mathcal{T}_y$ is intact for zero magnetic field, implying orthogonal symmetry. However, in the limit of vanishing ac scattering, $\Td/T\subac \rightarrow 0$, the Hamiltonian is block diagonal and $\mathcal{T}_y$ is ineffective since it connects states from different valleys. Within each valley the effective TRS associated with $\mathcal{T}_0$ is broken by the zz and im boundary conditions thus $\langle g_L \rangle$ is suppressed but never becomes positive. This is in contrast to the case of disordered bulk samples with small intervalley scattering rate, where the symmetry corresponding to $\mathcal{T}_0$ is not broken leading to weak anti-localization \cite{McCann2006}. That means we have two (nearly) decoupled unitary subsystems that lead to vanishing (or strongly suppressed) WL.

We point out that the magnetoconductance curves described by Eq.\,(\ref{eq:gL_weak_ac}) are squared Lorentzians and thus steeper and more narrow than Lorentzians with the same height (cf. dashed blue line in Fig.\,\ref{Fig:gL}). That means the magnetic dephasing is essentially assisted by the partial TRS breaking. \\

In Fig.\,\ref{Fig:gL} we summarize our findings graphically. The main panel shows the dependence of the WL correction on the magnetic flux, or $\sqrt{\Td/\TB} = \sqrt{\Td/\zeta}\,\Phi/\Phi_0$, respectively, for different values of the ratio $\Td/T\subac$.  As the ac scattering becomes stronger, the magnetoconductance gets closer to the Lorentzian behavior known from usual 2DEGs. The inset shows the crossover as a function of $\Td/T\subac$ for the case of zero magnetic flux. \\

To obtain Eq.\,(\ref{eq:gL}), we have assumed that the Ehrenfest time
\begin{equation}
 T\subt{E} = \frac{1}{\lambda} \ln(\vF \kE / \lambda)
\end{equation}
is much smaller than the dwell time $\Td$ and thus neglected terms of order $T\subt{E}/\Td$ or higher. $T\subt{E}$ is the time scale after which a minimal wave packet in a chaotic cavity has spread over the whole system, so that quantum interference is possible\cite{Zaslavsky1981}. For Schr\"odinger systems it is known\cite{Aleiner1996} that a finite Ehrenfest time leads to an exponential suppression of the WL correction. Since for graphene the ac scattering enters the integral (\ref{eq:modLoopdens_g}) in a similar way as the magnetic dephasing time, the semiclassical calculation\cite{Adagideli2003} is unchanged. Thus performing the integrals in Eq.\,(\ref{eq:sigL_ints}) for finite $T\subt{E}/\Td$, the WL is suppressed in the same way also here, namely in Eq.\,(\ref{eq:gL}) we get an overall exponential suppression
\begin{eqnarray}
\label{eq:gL_TE}
\hspace*{-0.7cm}
\frac{ \langle g_L \rangle}{g_0} 
&\approx&   - \frac{M_a M_b}{(M_a+M_b)^2} \hspace*{-0.1cm}\left(\frac{e^{-T\subt{E}/\Td}}{1+\frac{\Td}{\TB}} \hspace*{-0.05cm}- \hspace*{-0.05cm}
 \frac{e^{-T\subt{E}/\Td}}{1+\frac{\Td}{\TB}+2\frac{\Td}{T\subac}}\right). 
\end{eqnarray}

\subsubsection{Comparison with numerics}

\begin{figure}
 \centering
 \includegraphics[width=0.48\textwidth]{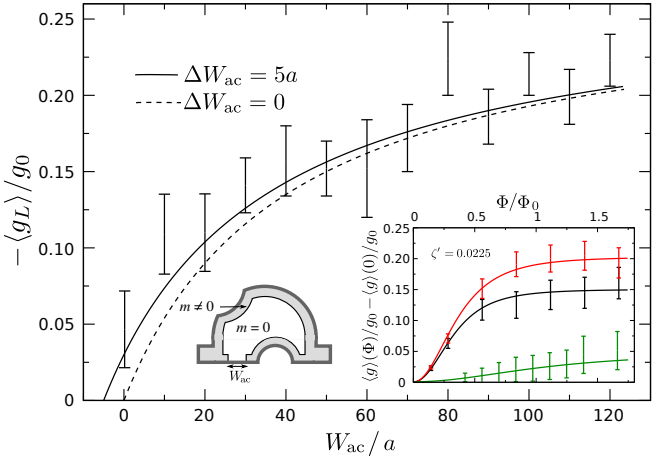}
\caption{\small{Comparison of the semiclassical theory for weak localization with numerical tight binding simulations. For the numerics we use a system (with leads $W_a=W_b=40\,a$) with a smooth mass edge modeling infinite mass boundaries\cite{Wurm2009} and open this mass boundary in order to tune $W_{\text{ac}}$. Fitting Eq.\,(\ref{eq:gL}) to the numerically calculated magnetoconductance, we obtain the total height of the weak localization signal $[1+T\subac /(2\Td)]^{-1}$ as a function of $W_{\text{ac}}$. The inset shows (from bottom to top) three examples for \mbox{$W_{\text{ac}}=0\,\text{(green)}$}, $\,50\,a\,\text{(black)~and~}110\,a\,\text{(red)}$ . We find $\zeta= 0.17\,\Td$ and for the limiting value $\langle g_L \rangle (\Phi=0, T\subac=0) = 0.27\,g_0$ from a fit to the numerical results for a system without a mass boundary that shows very strong intervalley scattering. The solid (dashed) line in the main panel displays the crossover as a function of $W_{\text{ac}}$ obtained from the semiclassical theory, Eq.\,(\ref{eq:gL_zeroPhi}), including an offset $\Delta W\subac = 5\,a$ ($\Delta W\subac = 0$). The error bars are centered around the numerical data.
}}
\label{Fig:gL_TB}
\end{figure}

In Fig.\,\ref{Fig:gL_TB} we compare our analytical results with the conductance obtained from numerical tight binding simulations. Here we consider a graphene cavity with the shape of a desymmetrized half stadium giving rise to chaotic scattering\cite{Bunimovich1979, Baranger1993a}. We model infinite mass boundaries by adding a smooth mass term close to the edges of the system (for details see Ref. \onlinecite{Wurm2009}). In order to tune $T\subac$ we cut a hole into the smooth mass so that locally a small region with armchair edges is present. The system is shown schematically as an inset in Fig.\,\ref{Fig:gL_TB}. We then calculate the magnetoconductance numerically (the inset shows three examples) using an adaptive recursive Green's function method \cite{Wimmer2009} and obtain $\langle g_L \rangle$  as the difference of $\langle g \rangle (\Phi)$ at zero flux and $\Phi \approx 1.4-1.7 \Phi_0$. The main panel in Fig.\,\ref{Fig:gL_TB} compares the absolute value of the WL correction obtained in this way with the semiclassical prediction (\ref{eq:gL_zeroPhi}). We find that the numerical data reproduces our semiclassical theory very well (dashed line). The agreement is even better if we assume an effective offset in $W\subac$ of five lattice constants, $\Delta W\subac = 5 a$ (solid line). This offset can be explained by the fact that our smooth mass edges cannot completely avoid intervalley scattering from armchair edges or sharp edges at the boundary, e.\,g.\,at the lead mouths. We assign further deviations of the numerical from the anayltical results to the limitations of the Dirac equation and the semiclassical approximation. 

%

\subsection{Universal Conductance Fluctuations}
\begin{figure}
 \centering
 \includegraphics[width=0.48\textwidth]{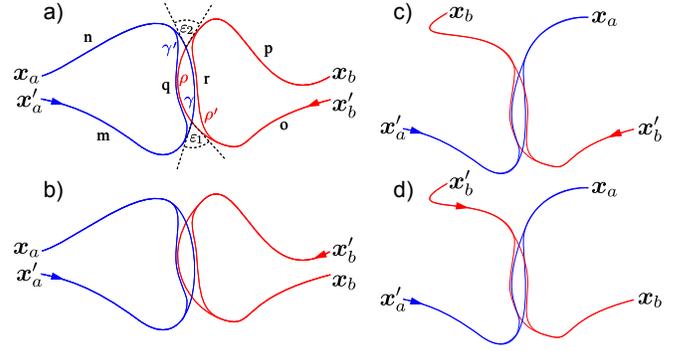}
\caption{\small{Schematic view of orbit quadruplets that give rise to the conductance fluctuations in leading order, see Ref.\,\onlinecite{Brouwer2006}. Here we have two (avoided) crossings of orbits from different leads. The pieces of the orbits before, after and between the crossings are labeled with letters $n$ to $r$.}}
\label{Fig:squids}
\end{figure}
So far we have considered quantum effects in the averaged conductance, i.\,e.\,the smooth part of $g$. Now we address quantum fluctuations around this average value, often denoted as universal conductance fluctuations (UCF) because their variance does not scale with the average conductance of a disordered or chaotic system. 
The UCF in diffusive graphene samples have been investigated in several theoretical\cite{Kharitonov2008, Kechedzhi2008} and experimental\cite{Morozov2006, Lundeberg2009, Horsell2009, Berezovsky2010a} studies. Here we consider the case of ballistic graphene cavities.
To quantify the size of these fluctuations, we calculate the variance of the conductance $g_{ab}$ which is identical to the covariance
of the resistances $g_{aa}$ and $g_{bb}$:
\begin{eqnarray}
\label{eq:vardef}
 \text{var}(g_{ab}) 
 = \langle g_{aa}g_{bb} \rangle - \langle g_{aa}\rangle\langle g_{bb} \rangle\,.
\end{eqnarray}
Here the semiclassical\cite{Baranger1993a, Brouwer2006} and RMT\cite{Beenakker1997, Baranger1994, Jalabert1994} prediction for many propagating lead channels reads
\begin{eqnarray}
\frac{\text{var}(g_{ab})}{g_0^2} 
=\frac{2M_aM_b}{\beta(M_a+M_b)^4} + \mathcal{O}\left(M_{a/b}^{-1}\right) 
\end{eqnarray}
for a scalar Schr\"odinger equation.
We consider the first term in Eq.\,(\ref{eq:vardef}), the averaged product of resistances.
According to Eq.\,(\ref{eq:sigma_sc}), multiplication of two resistivities results in the fourfold sum over classical orbits
\begin{eqnarray}
 &&\hspace*{-0.3cm}\sigma_{aa} \sigma_{bb} = \left(\frac{ e^2 \vF^2}{16\pi^2\hbar^2} \right)^2 \sum_{\gamma \gamma'(a)} \sum_{\rho \rho'(b)}  X_{\gamma,\gamma',\rho,\rho'}
 D_\gamma D_{\gamma'}D_\rho D_{\rho'}\nonumber \\
 &\times& \exp\left[i\delta S_{\gamma,\gamma'}/\hbar +i\delta S_{\rho,\rho'}/\hbar+i\delta\mu_{\gamma,\gamma'}{\pi}/{2}+i\delta\mu_{\rho,\rho'}{\pi}/{2}\right]
\nonumber \,.
\end{eqnarray}
with
\begin{equation}
\label{eq:Xdef}
 X_{\gamma,\gamma',\rho,\rho'} =  \tr\left(\sigmaa \Ktilde_{\gamma}\sigmaa  \Ktilde^\dagger_{\gamma'}\right)
 \tr\left(\sigmab \Ktilde_{\rho}\sigmab  \Ktilde^\dagger_{\rho'}\right)\,.
\end{equation}
Note that the orbits $\gamma$ and $\gamma'$ originate from the same point $\bsy{x}_a'$ in lead $a$ and end at the same point $\bsy{x}_a$ also in lead $a$.
The orbits $\rho$ and $\rho'$ on the other hand both begin at $\bsy{x}_b'$ in lead $b$ and end at $\bsy{x}_b$ in lead $b$.
Here the averaging procedure gives only a significant result when the \textit{sum of the action differences} is small or vanishing.
One possibility is that the action differences of the orbits from one lead, i.\,e.\,$\delta S_{\gamma,\gamma'}$ and $\delta S_{\rho,\rho'}$, are both small individually
and the orbits from different leads are uncorrelated. These contributions are fully contained in the second term, namely the product of averages, and 
thus they cancel out. Therefore we have to consider orbit quadruplets in which only the complete phase is small, but not the individual phases. 
Neglecting effects due to a finite Ehrenfest time, the dominant pairings are those sketched in Fig.\,\ref{Fig:squids}, as shown by Brouwer and Rahav\cite{Brouwer2006}. 
Two orbits from different leads, say $\gamma$ and $\rho$, cross each other with a crossing angle $\varepsilon_1$ and then cross each other again, now with a crossing angle $\varepsilon_2$. The two other orbits follow the first ones exponentially closely, but avoid the crossings. For instance $\gamma'$ follows $\gamma$ to the first crossing, then it follows $\rho$ to the second crossing and finally again $\gamma$ back to the lead [Fig.\,\ref{Fig:squids}\,a) and b)].
We label the individual pieces with $m,n,o,p,q,r$ as marked in Fig.\,\ref{Fig:squids}\,a) and denote the combination of the pieces $r$ and $q^{-1}$, which is $q$ traversed in opposite direction, by the loop part $L$. Note that also quadruplets where two orbits from different leads cross first and then avoid the second crossing (or vice versa) while the two other orbits avoid the first crossing and cross then (or vice versa) give rise to a contribution to the UCF in leading order [Fig.\,\ref{Fig:squids}\,c) and d)]. 

For the details of the calculation of $X_{\gamma,\gamma',\rho,\rho'}$ we refer to App.\,\ref{app:pspintrace_5}. We note however that, in contrast to the individual traces in Eq\,(\ref{eq:Xdef}), their product $X_{\gamma,\gamma',\rho,\rho'}$ depends only on the reflections along the loop part $L$. In fact we find $X_{\gamma,\gamma',\rho,\rho'}=0$ if the total number of reflections from ac edges during the loop part is odd, and for an even number of ac reflections we get 
\begin{eqnarray}
\label{eq:X_ggrr}
X_{\gamma,\gamma',\rho,\rho'} &=& 64 \cos(\theta_a) \cos(\theta_a')\cos(\theta_b) \cos(\theta_b') \nonumber \\
&&\times 
\cos^2(2K\Lambda_L + \vartheta_L + N_{{L}}\pi/2)\,.
\end{eqnarray}
In order to perform the orbit sums, we make use of the fact that the two self-intersections are independent and thus their contributions factorize\cite{Heusler2006, Brouwer2006}
so that we get
\begin{eqnarray}
\label{eq:varint}
  \text{var}(\sigma_{ab}) \hspace*{-0.26cm}&&\approx \left(\frac{2e^2 \vF \kE W_a W_b}{\pi^3\hbar A}\right)^2 \\
 &\times& \mathrm{Re} \intl_{0}^\pi d\varepsilon_1\!\!\!\! \intl_{2T_{\text{min}}(\varepsilon_1)}^\infty \!\!\!\!dT_1\, 
e^{-[T_1-T_{\text{min}}(\varepsilon_1)]/\Td} e^{i \delta S(\varepsilon_1)/\hbar}\nonumber \\
 &\times& \mathrm{Re} \intl_{0}^\pi d\varepsilon_2\!\!\!\! \intl_{2T_{\text{min}}(\varepsilon_2)}^\infty \!\!\!\!dT_2\, 
e^{-[T_2-T_{\text{min}}(\varepsilon_2)]/\Td} e^{i \delta S(\varepsilon_2)/\hbar} \nonumber \\
&\times&\breve{P}_{\text{ac}}(\varepsilon_1,\varepsilon_2,T_1,T_2) \,. \nonumber 
\end{eqnarray}
Here $\breve{P}_{\text{ac}}(\varepsilon_1,\varepsilon_2,T_1,T_2)$ is the density of two uncorrelated crossings such
that armchair pieces are hit either an even number of times during both the $q$ and the $r$ part of the loop or 
an odd number of times during both parts
\begin{eqnarray}
\label{eq:Pbreve}
 \breve{P}\subac (\varepsilon_1,\varepsilon_2,T_1,T_2)  &\approx &
 \left( \frac{\vF^2 \varepsilon}{\pi A} \right)^2 \intl_{T_{\text{min}}(\varepsilon_1)}^{T_1-T_{\text{min}}(\varepsilon_1)} \hspace*{-0.5cm} dT_q 
 \intl_{T_{\text{min}}(\varepsilon_2)}^{T_2-T_{\text{min}}(\varepsilon_2)}  \hspace*{-0.5cm} dT_r \nonumber \\
&& \hspace*{-2.5cm} \times\,  [T_1-T_q -T_{\text{min}}(\varepsilon_1)]
 [T_2-T_r -T_{\text{min}}(\varepsilon_2)] \nonumber \\
 && \hspace*{-2.5cm} \times \,  P\supt{e}\subac[T_q + T_r - T_{\text{min}}(\varepsilon_1) - T_{\text{min}}(\varepsilon_2)]\,.
\end{eqnarray}
To obtain Eq.\,(\ref{eq:varint}) we have used the sum rule (\ref{eq:sumrule}) twice, included the averaged pseudospin traces $\langle X_{\gamma,\gamma',\rho,\rho'} \rangle$ for the contributing quadruplets and integrated over the incoming and outgoing angles as well as over the lead openings. 
Furthermore we included a factor of two due to time reversed quadruplets, i.\,e.\,such quadruplets, where the orbits from different leads (e.\,g.\,$\gamma$ and $\rho'$) traverse the corresponding part of the loop ($q$) in opposite direction rather than in parallel [panel b) vs. a) and d) vs. c) in Fig.\,\ref{Fig:squids}].
For zero magnetic field, both contribute equally, and analog as for Eq.\,(\ref{eq:gL}) we get for the variance of the conductance\cite{Richter2002, Waltner2010}
\begin{eqnarray}
\label{eq:var_zeroPhi}
\frac{\text{var}(g_{ab})}{g_0^2} 
\approx \frac{2M_a^2M_b^2}{(M_a+M_b)^4} \left[1+ \frac{1}{\left(1+2\Td /T\subac\right)^2}\right] \,,
\end{eqnarray}
which is the central result of this section.
The crossover described by this equation is different from the one for the average conductance, Eq.\,(\ref{eq:gL_zeroPhi}). Also here we have no explicit dependence on $\kE$, but as for the spectral form factor, degeneracies have to be taken into account when interpreting the result. For identical leads we have a crossover from $1/4$ to $1/8$ for an increasing probability of armchair scattering. This agrees with a transition from two degenerate unitary subsystems ($4\times$CUE) to one orthogonal system (COE). 

\begin{figure}
 \centering
 \includegraphics[width=0.48\textwidth]{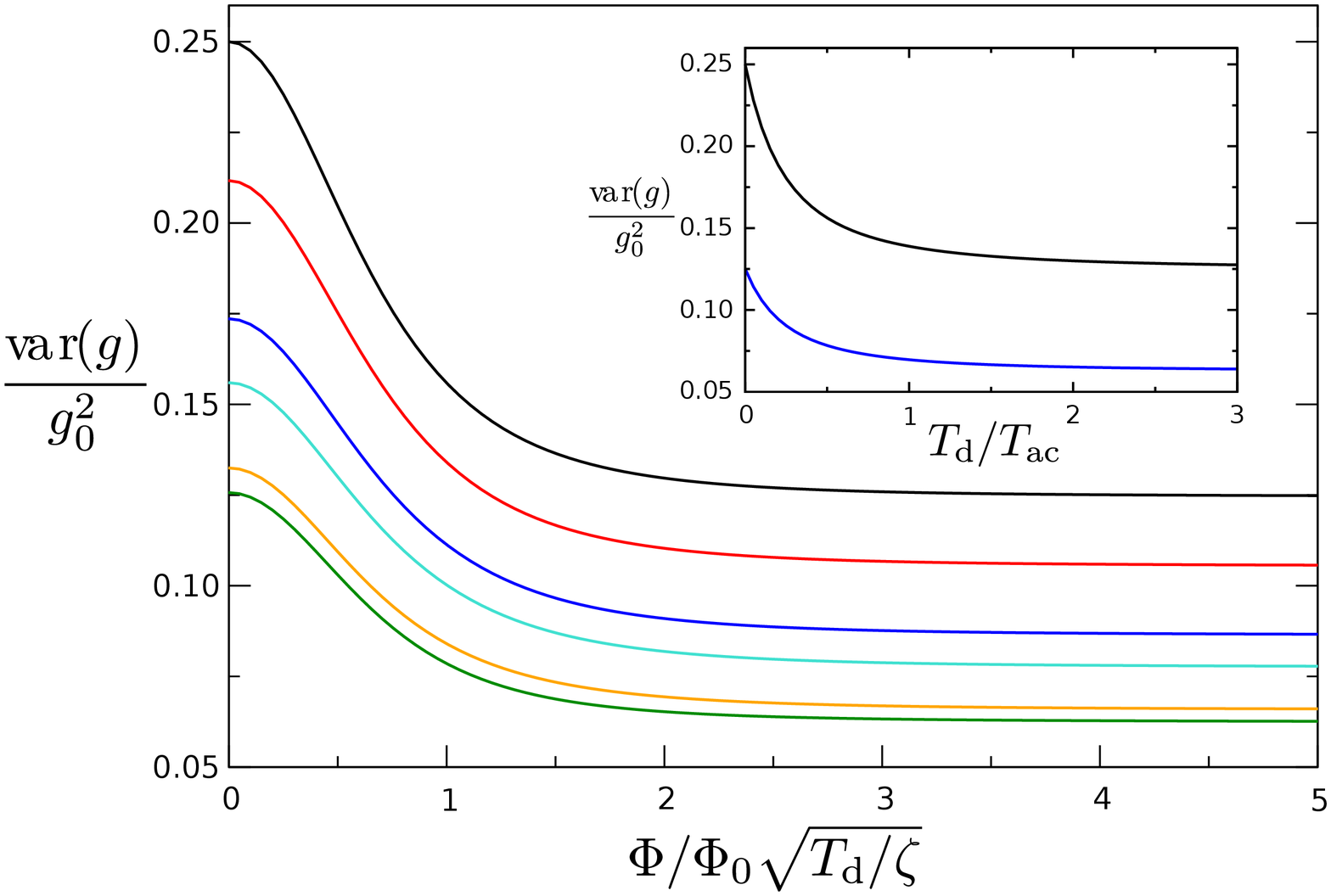}
\caption{\small{Variance of the conductance as a function of the magnetic flux $\Phi$ through the system obtained 
from our semiclassical theory for identical leads.
The curves correspond to different values of the relative armchair scattering strength $\Td /T\subac$ (from top to bottom): 0.0 (black), 0.1 (red), 0.3 (blue), 0.5 (turquoise), 1.5 (orange) and
5 (green).
The inset shows the crossover as a function of $\Td /T\subac$ at zero magnetic flux $\Phi=0$ (upper black curve) and at finite flux $\Phi=5\,\Phi_0/\sqrt{\Td/\zeta}$ (lower blue curve) [cf. Eq.\,(\ref{eq:var_zeroPhi}) and Eq.\,(\ref{eq:var})].
}}
\label{Fig:var}
\end{figure}

For a finite magnetic flux through the cavity we have to modify the densities accordingly as in the previous sections. However, not all quadruplets are affected by
the magnetic field, because there is an additional phase due to the flux only if the loop part is traversed in opposite directions by the different orbits.
Thus we get one half of the zero flux result plus the contribution from the quadruplets that are affected by the magnetic dephasing for the variance:
\begin{equation}
\label{eq:var}
 \frac{\text{var}(g_{ab})}{g_0^2}  
 = \frac{M_a^2M_b^2}{(M_a+M_b)^4} \sum_{i,j = 0,1} 
 \frac{1}{\left(1+i\,\Td /\TB+2j\,\Td /T\subac \right)^2}\,.
\end{equation}
Even though \textit{all orbit quadruplets} are affected by the `dephasing' due to the armchair scattering, the magnetic flux has an effect only on 
\textit{one half of the orbit quadruplets} and both mechanisms lead to a similar behavior of the variance in the final result (\ref{eq:var}).
For identical leads and $T\subac \rightarrow \infty$, Eq.\,(\ref{eq:var}) describes a crossover from $1/4$ to $1/8$ with increasing flux $\Phi$, i.\,e.\,
from two degenerate unitary subsystems ($4\times$CUE) to two independent unitary subsystems ($2\times$CUE), and for 
$T\subac \rightarrow 0$ from $1/8$ to $1/16$, i.\,e.\,from an orthogonal (COE) to a unitary (CUE) system, in agreement with our
findings for the average conductance and the spectral form factor.
In Fig.\,\ref{Fig:var} we depict our semiclassical result (\ref{eq:var}). We show the magnetodependence of the conductance variance for different values
of the ac scattering time $T\subac$ (main panel) as well as the dependence of $\text{var}(g)$ on the ratio $\Td/T\subac$ for zero (inset, upper black curve)
and finite (inset, lower blue curve) magnetic flux. The crossover as a function of both the ac scattering and the magnetic flux is here faster than for the
WL correction to the average conductance, because the denominators in Eq.\,(\ref{eq:var}) are squared as opposed to those in Eq.\,(\ref{eq:gL}).

\begin{figure}
 \centering
 \includegraphics[width=0.48\textwidth]{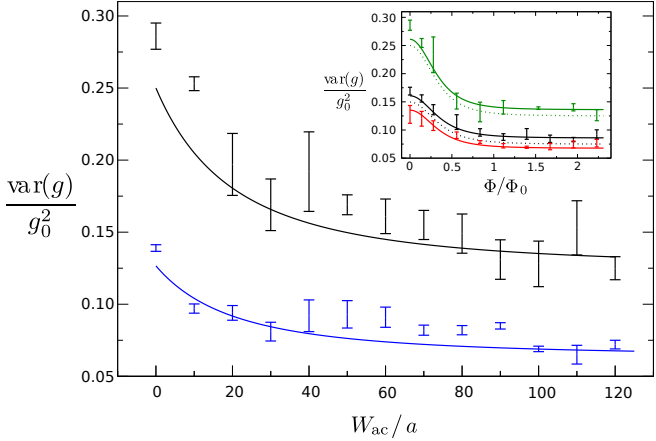}
\caption{\small{Comparison of our semiclassical theory, Eq.\,(\ref{eq:var}), with numerical tight binding simulations. For the numerics we used the same system as for the average conductance. 
The main panel shows the crossover as a function of $W_{\text{ac}}$ obtained from the semiclassical theory Eq.\,(\ref{eq:var}) (solid line) and from  the numerical data (lines with error bars) at zero magnetic flux $\Phi=0$ (upper black curve) and at finite flux $\Phi=1.4\, \Phi_0$ (lower blue curve).
The inset shows (from top to bottom) the dependence of the numerical data on the magnetic flux (lines with error bars) for systems with $W_{\text{ac}}=0\,$ (green), $50\,a\,$ (black) and $110\,a\,$ (red). We find that $\zeta \approx 0.25\,\Td$ fits the data well. For the solid lines at $W_{\text{ac}}=0$ and $50\,a$, we 
added a constant offset of $0.011$ to the result of Eq.\,(\ref{eq:var}), the dotted lines show the result without any offset.
}}
\label{Fig:var_TB}
\end{figure}

Fig.\,\ref{Fig:var_TB} compares the semiclassical result with numerical tight binding calculations. The system used for the numerical study is the same as we used for 
the average conductance. We find that Eq.\,(\ref{eq:var}) provides a good approximation to the numerical data.
%


\subsection{Shot noise}
\begin{figure}
 \centering
 \includegraphics[width=0.38\textwidth]{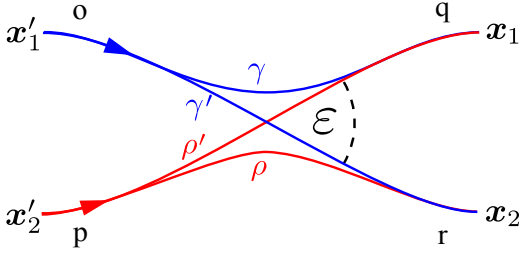}
\caption{\small{Sketch of the orbit quadruplets contributing to the shot noise in leading order. As opposed to the UCF only one crossing is present.
Orbits that begin at the same point, end in general at different points.}}
\label{Fig:ShotNDia}
\end{figure}

As another important transport property, we briefly discuss the shot noise of a ballistic graphene cavity.
For graphene ribbons with rather large width to length ratio, where transport at very low energies may be dominated by evanescent modes, shot noise has been studied theoretically\cite{Tworzydlo2006, Cheianov2006} and experimentally\cite{Danneau2008, DiCarlo2008}. We focus here on the case of chaotic scattering at larger energies, applying our semiclassical theory.
Usually the shot noise is quantified via the \textit{Fano factor} $\mathcal{F}$, defined as the ratio of the \textit{shot noise power} $S$
of the system and the Poissonian shot noise power $S\subt{P}$ corresponding to charge carriers being transmitted in an uncorrelated way from
one lead to the other. In terms of the transmission amplitudes $t = t_{ab}$ the Fano factor, that describes the quantum suppression of the shot noise with respect to the classical value, is given by\cite{Buttiker1990}
\begin{equation}
 \mathcal{F} = \frac{S}{S\subt{P}} = \frac{\tr\left(tt^\dagger - tt^\dagger tt^\dagger\right)}{ \tr\left(tt^\dagger \right)}\,.
\end{equation}
Using RMT for the scattering matrix one obtains\cite{Jalabert1994}
\begin{equation}
\label{eq:Fano_RMT}
\mathcal{F} = \frac{M_aM_b}{(M_a+M_b)^2} + \mathcal{O}(M_{a/b}^{-1})\,.
\end{equation}
In App.\,\ref{app:FisherLee}, where we discuss the generalized Fisher-Lee relations, we show that
\begin{eqnarray}
 \hspace*{-0.78cm} \tr\left(tt^\dagger \right) &=& - \frac{h}{e^2} \int\limits_{C_a}dy\int\limits_{C_b}dy'~ \sigma_{ab}(\bsy{x},\bsy{x'})\,, \\
\label{eq:ShotN_tttt}
 \hspace*{-0.78cm} \tr\left(tt^\dagger tt^\dagger\right) &=& \int\limits_{C_a} \!\! dy_1\!\! 
 \int\limits_{C_a}\!\! dy_2\!\! \int\limits_{C_b}\!\! dy'_1\!\! \int\limits_{C_b}\!\! dy'_2~ \tilde{\sigma}_{ab}(\bsy{x}_1,\bsy{x}_2,\bsy{x}_1',\bsy{x}_2') 
\end{eqnarray}
with 
\begin{eqnarray}
\label{eq:ShotN_sigma}
 \hspace*{-0.1cm} && \tilde{\sigma}_{ab}(\bsy{x}_1,\bsy{x}_2,\bsy{x}_1',\bsy{x}_2') \\  
 \hspace*{-0.1cm}  &=&\hspace*{-0.05cm} \tr\hspace*{-0.1cm}\left[\sigmaa G(\bsy{x}_1,\bsy{x}_1') \sigmab G^{\dagger}\hspace*{-0.05cm}(\bsy{x}_2,\bsy{x}_1')
 \sigmaa G(\bsy{x}_2,\bsy{x}_2') \sigmab G^{\dagger}\hspace*{-0.05cm}(\bsy{x}_1,\bsy{x}_2')\right] \,. \nonumber
\end{eqnarray}
Inserting the semiclassical Green's function (\ref{eq:Gsc}) we get the fourfold sum 
\begin{eqnarray}
&& \tilde{\sigma}_{ab}(\bsy{x}_1,\bsy{x}_2,\bsy{x}_1',\bsy{x}_2') =  \left(\frac{ \vF^2}{8\pi\hbar} \right)^2\\
&\times&  \sum_{\gamma \gamma'} \sum_{\rho \rho'} \tr\left(\sigmaa \Ktilde_{\gamma}\sigmab  \Ktilde^\dagger_{\gamma'} \sigmaa \Ktilde_{\rho}\sigmab  \Ktilde^\dagger_{\rho'}\right) D_\gamma D_{\gamma'}D_\rho D_{\rho'} \nonumber \\ 
&\times& \exp\left[i\delta S_{\gamma,\gamma'}/\hbar+i \delta S_{\rho,\rho'}/\hbar+i\delta\mu_{\gamma,\gamma'}{\pi}/{2}+i\delta\mu_{\rho,\rho'}{\pi}/{2}\right]
\nonumber \,,
\end{eqnarray}
similar as for the UCF.
However, for the UCF the contributing classical orbits came in pairs with the same starting \textit{and} end points. 
Here, the orbits with the same starting point have in general
different end points and vice versa. In words, the trajectory $\gamma$ starts at a point $\bsy{x}_1'$ in lead $b$ and ends at a point $\bsy{x}_1$ in lead $a$, the trajectory $\gamma'$ starts also at $\bsy{x}_1'$ but ends at a point $\bsy{x}_2$ in lead $a$. The trajectories $\rho$ and $\rho'$ both begin at a point $\bsy{x}_2'$ in lead $b$
and $\rho$ ends at $\bsy{x}_2$ while $\rho'$ ends at $\bsy{x}_1$.
The diagonal contribution corresponds to quadruplets where either $\gamma=\gamma'$ and $\rho=\rho'$ or $\gamma=\rho'$ and $\rho=\gamma'$.
This is only possible if either the end points are equal $\bsy{x}_1=\bsy{x}_2$ or the starting points $\bsy{x}_1'=\bsy{x}_2'$. One can show that
in this case $\tr\left(tt^\dagger \right) = \tr\left(tt^\dagger tt^\dagger\right)$\cite{Lassl2003, Schanz2003}, and the Fano factor $\mathcal{F}$ is zero in diagonal approximation. This is the classical limit, where no quantum shot noise exists.
The leading order contribution to $\mathcal{F}$ is known to stem from diagrams as sketched in Fig.\,\ref{Fig:ShotNDia} \cite{Lassl2003, Muller2007, Schanz2003}.
For these quadruplets the valley part of the pseudospin trace does not depend on the edge type at all, since
\begin{eqnarray}
\tr\left(\Ktilde^{\text{v}}_{\gamma} \Ktilde^{\text{v}\dagger}_{\gamma'}\Ktilde^{\text{v}}_{\rho} \Ktilde^{\text{v}\dagger}_{\rho'}\right)  
&=& \tr\left(\Ktilde^{\text{v}}_{q} \Ktilde^{\text{v}}_{o}\Ktilde^{\text{v}\dagger}_{o} \Ktilde^{\text{v}\dagger}_{r}\Ktilde^{\text{v}}_{r} \Ktilde^{\text{v}}_{p} \Ktilde^{\text{v}\dagger}_{p} \Ktilde^{\text{v}\dagger}_{q}\right) \nonumber \\ &=& 2 \,.
\end{eqnarray}
Also for the sublattice part, we do not find such an edge dependence. A longer calculation similar to the one for the average conductance and the UCF results in
\begin{eqnarray}
&& \tr\left(\sigmaa\Ktilde^{\text{s}}_{\gamma}\sigmab \Ktilde^{\text{s}\dagger}_{\gamma'}\sigmaa\Ktilde^{\text{s}}_{\rho} \sigmab\Ktilde^{\text{s}\dagger}_{\rho'}\right)
\nonumber \\
&=& 32 \cos(\theta_1)\cos(\theta_1')\cos(\theta_2)\cos(\theta_2') \,.
\end{eqnarray}
Hence to leading order in the inverse channel number the Fano factor of chaotic graphene flakes does not depend on the edge characteristics. The orbit sums can be performed exactly as in the
case of usual Schr\"odinger systems giving
\begin{equation}
 \tr\left(tt^\dagger - tt^\dagger tt^\dagger\right) \approx \frac{M_a^2M_b^2}{(M_a+M_b)^3}
\end{equation}
which, together with Eq.\,(\ref{eq:gD}), yields the leading order term of Eq.\,(\ref{eq:Fano_RMT}). To conclude, at least to leading order in inverse channel number, the shot noise does not exhibit any imprints of the edges of a chaotic graphene cavity.

\section{Conclusions}
\label{sec:conclusions}

The major part of theoretical work on quantum interference phenomena in mesoscopic graphene systems has focused on the diffusive regime. 
However due to the permanent progress in manufacturing techniques
it becomes possible to fabricate clean structures, where the main scattering mechanism is due to reflection from the system's edges. 
In this work, we provide a comprehensive semiclassical study of quantum interference effects in ballistic graphene nanostructures, particularly focusing on the effects 
of the edge characteristics. 
To this end we have started from a general analytic expression for the full Green's function of a graphene cavity with arbitrary edge structure\cite{PartI} and
generalized semiclassical methods that have been originally developed for Schr\"odinger systems to graphene nanostructures.

As a first application we have have studied the fluctuations of the density of states of closed graphene billiards with chaotic classical dynamics. 
Since the DOS of such billiards is very complex, we have considered its statistical properties, namely the two-point correlations of the DOS via the spectral form factor $F(t)$. Already in the diagonal approximation we find a strong dependence of $F$ on the specific edge structure. We find that the total amount of intervalley scattering armchair boundaries sets the timescale for a crossover in the form factor, where the asymptotic limits agree with the RMT predictions for two degenerate unitary subsystems ($4\times$GUE) in the absence of intervalley coupling and a single orthogonal system (GOE) in the limit of complete valley mixing respectively. We also have derived the explicit edge dependence of the simplest off-diagonal corrections, confirming our interpretations of the diagonal contributions. Even for a significant amount of intervalley scatting ac edges, we predict that the small $t$ correlations deviate significantly from the GOE result. Our main results for the DOS, Eqs.\,(\ref{eq:FD_gen}) and (\ref{eq:FL_gen}), thus suggest that in experiments, the effects of effectively broken TRS should be visible in the correlations of the DOS on a scale of many mean level spacings rather than in the usually considered next nearest neighbor statistics, where intervalley coupling is essentially too effective to see those effects\cite{Wurm2009}. 
Our results do not support the experimental indication of GUE statistics in the nearest neighbor level distribution in Ref.\,\onlinecite{Ponomarenko2008}. The latter would suggest extremely weak intervalley scattering or some other, yet unknown, symmetry breaking mechanism.

In the second part of the paper, we have further derived analytical expressions for the most important transport properties of open ballistic graphene cavities in the semiclassical limit, such as the average conductance, the universal conductance fluctuations and the Fano factor.
We have calculated the classical value (diagonal approximation) and the leading order quantum corrections, i.\,e.\,the weak localization correction, for the average conductance.
While the classical conductance does not depend on the edge structure of the cavity, the WL correction is again governed by the effective armchair scattering time. 
Equation (\ref{eq:gL}) gives the WL correction of a graphene flake in terms of its microscopic edge properties. The WL is suppressed if intervalley coupling is absent, i.\,e.\,there is no weak antilocalization in this case, because the effective intravalley TRS is not preserved in the presence of edges. With increasing ac scattering strength, the intervalley TRS becomes effective again and the WL correction reaches the COE value for very strong intervalley coupling. For the UCF we find corresponding crossovers between the unitary and the orthogonal random matrix ensembles in agreement with our findings for the WL and the spectral form factor. 
Finally we have shown that the shot noise power and thus the Fano factor are not affected by the edge structure to leading order.
Our main results for the transport properties, Eqs.\,(\ref{eq:gL}) and (\ref{eq:var}) explain part of our earlier numerical work\cite{Wurm2009}, and provide a theoretical footing for experimental and numerical studies of graphene samples in the ballistic regime.

\section*{Acknowledgments}

We thank Jack Kuipers, Adam Rycerz, Daniel Waltner and Michael Wimmer for helpful discussions.
This work was supported by the DFG within GRK 1570 (JW, KR) and through TUBA under 
grant I.A/TUBA-GEBIP/2010-1 as well as the funds of the Erdal \.{I}n\"on\"u chair at Sabanc\i~University (IA).
JW further acknowledges hospitality from Sabanc\i~University.


\appendix

\section{Calculation of pseudospin traces}

\subsection{Pseudospin trace for the form factor $F(t)$}
\label{app:pspintrace_1}
We calculate the trace of ${K}_\gamma$, Eq.\,(\ref{eq:trace-K}), for a given classical orbit $\gamma$ taking into account reflections from edges of infinite mass, armchair, and zigzag type. We separate the pseudospin propagator into a sublattice and a valley part, ${K}_\gamma = {K}_\gamma^{\text{v}} \otimes {K}_\gamma^{\text{s}}$, and consider first the valley part ${K}_\gamma^{\text{v}}$. It is traceless if the number of ac reflections is odd, and otherwise we have [cf. Eq.\,(\ref{eq:K_i})]
\begin{equation}
  {K}_\gamma^{\text{v}} = \pm\tau_z^N e^{i(2K \Lambda_\gamma +\vartheta_\gamma)\tau_z} 
\end{equation}
and thus
\begin{equation}
 \tr {K}_\gamma^{\text{v}} = \pm i^N 2 \cos(2K\Lambda_\gamma + \vartheta_\gamma- N\pi/2)\,,
\end{equation}
with $\Lambda_\gamma$ and $\vartheta_\gamma$ as defined in Eqs.\,(\ref{eq:Lambda_def}) and (\ref{eq:vartheta_def}). The sign depends on the exact sequence
of zz, ac, and im reflections and the individual signs in  Eq.\,(\ref{eq:K_i}).
Every pair of successive ac reflections gives rise to
\begin{equation}
 \tau_y e^{2iK x_{i}} \tau_y e^{2iK x_{i-1}} = e^{2iK (x_{i-1} - x_i)}
\end{equation}
and thus to $\Lambda_\gamma$.\\
For the sublattice part, we begin assuming that only one reflection along $\gamma$ is at a zz edge, without loss of generality the very first one, and afterwards an even number of reflections occur at ac edges and the rest at im edges. Then we can cast ${K}_\gamma^{\text{s}}$ in the form
\begin{eqnarray}
{K}_\gamma^{\text{s}} &=& i^N \sigma_z^{N-1} e^{i (\theta_\gamma - \theta_1) \sigma_z} \sigma_{\bsy{t}_1} (1+\sigma_{\bsy{\alpha}_1,\bsy{\alpha}_N} ) 
\end{eqnarray}
and therefore
\begin{equation}
\tr{K}_\gamma^{\text{s}}
\!= \! 2 i^{2N-1}\! \cos\left[ \theta_\gamma \!\!-\! \theta_1\! + \!\varphi_{\bsy{\alpha}_1,\bsy{\alpha}_N}\!\! -\! \varphi_{\bsy{t}_1} \!-\! (N\!-\!1)\pi/2 \right],
\end{equation}
where $\varphi_{\bsy{v}}$ is the polar angle of the vector $\bsy{v}$.
Geometric considerations give further
\begin{equation}
\label{eq:}
 \varphi_{\bsy{\alpha}_1,\bsy{\alpha}_N} - \varphi_{\bsy{t}_1} = \theta_1 - \pi/2 \mod 2\pi
\end{equation}
and thus we get
\begin{equation}
\label{eq:sltrace}
 \tr{K}_\gamma^{\text{s}} = 2 i^{2N-1} \cos\left( \theta_\gamma - N\pi/2 \right)\,.
\end{equation}
Similarly we write for the case of two zz reflections
\begin{equation}
 {K}_\gamma^{\text{s}} = i^N \sigma_z^{N^{(2)}} e^{i \theta^{(2)}_\gamma \sigma_z} \sigma_{\bsy{t}_2} \sigma_z^{N^{(1)}} e^{i \theta^{(1)}_\gamma \sigma_z} \sigma_{\bsy{t}_1} (1+\sigma_{\bsy{\alpha}_1,\bsy{\alpha}_N} ) 
\end{equation}
leading to 
\begin{eqnarray}
\tr{K}_\gamma^{\text{s}}
&=& 2  i^{2N+2} (-1)^{N^{(1)}+1} \nonumber \\ 
&\times& \cos\left(\theta^{(2)}_\gamma - \theta^{(1)}_\gamma + \varphi_{\bsy{t}_1} - \varphi_{\bsy{t}_2} - N \pi/2 \right) \,.
\end{eqnarray}
Once more employing a geometric relation
\begin{equation}
\theta^{(2)}_\gamma - \theta^{(1)}_\gamma + \varphi_{\bsy{t}_1} - \varphi_{\bsy{t}_2} = \theta_\gamma - (N^{(1)}+1 ) \pi \mod 2\pi
\end{equation}
we find
\begin{equation}
 \tr{K}_\gamma^{\text{s}} = 2 i^{2N-2} \cos\left( \theta_\gamma - N\pi/2 \right)\,.
\end{equation}
The treatment of these two special cases can be combined to find the pseudospin trace for an arbitrary number of zz reflections. Absorbing all phases in the prefactors into $f_\gamma$, we finally obtain Eq.\,(\ref{eq:trace-K}).

\subsection{Trace $Z_{\gamma,\gamma_\times}$ for the loop contribution to $F(t)$}
\label{app:pspintrace_2}

From the discussion above, we know already that 
$\tr \,\Ktilde_\gamma = \tr\, \Ktilde_{\gamma_\times}=0 $ if the total number of armchair reflections is odd, so we can concentrate on the opposite case.
By using Eq.\,(\ref{eq:sep_theta}), we find 
\begin{eqnarray}
 &&\cos(\theta_\gamma+N\pi/2)\cos(\theta_{\gamma_\times}+N\pi/2) \nonumber \\ &=&
 \cos^2(\theta_L+N\pi/2)\cos^2(\theta_R) - \sin^2(\theta_L+N\pi/2)\sin^2(\theta_R) \nonumber \\ &=& (-1)^{N_R+1}+ \mathcal{O}(\varepsilon^2)\,,
\end{eqnarray}
where we made use of the expressions equivalent to Eq.\,(\ref{eq:thetasplit})
\begin{eqnarray}
\theta_L = (N_{+,L} - N_{-,L} +1)\frac{\pi}{2} -w_L\pi + \mathcal{O}(\varepsilon)\,, \\
\theta_R = (N_{+,R} - N_{-,R}-1)\frac{\pi}{2}  -w_R\pi  + \mathcal{O}(\varepsilon) \,.
\end{eqnarray}
Since in the semiclassical limit the main contribution to $F_L$ comes from pairs with small angles $\varepsilon$\cite{Sieber2001}, we keep only the leading term $(-1)^{N_R+1}$. 
With that Eq.\,(\ref{eq:TrKgTrKgx}) for the loop pairs with even $N_{\text{ac}}$ follows from Eq.\,(\ref{eq:trace-K}).

\subsection{Trace $Y_\gamma$ for the diagonal contributions to the conductance}
\label{app:pspintrace_3}

For the sublattice part of $Y_\gamma$ we consider
orbits that hit armchair and infinite mass edges. Each reflection effectively contributes a term $\sigma_z e^{i\theta_i\sigma_z}$ and thus
\begin{eqnarray}
 &&\tr\left(\sigmaa \Ktilde_{\gamma}\sigmab  \Ktilde^\dagger_{\gamma}\right)\nonumber \\ &=&
2\, \tr\left[\sigmaa \sigma_z^N e^{i\theta_\gamma \sigma_z} (1+\sigmav{\bsy{\alpha}_1 \bsy{x'}}) 
 \sigmab (1+\sigmav{\bsy{\alpha}_1 \bsy{x'}}) e^{-i\theta_\gamma \sigma_z} \sigma_z^N \right] \nonumber \\
&=& (-1)^N 2 \tr\left[e^{-2i\theta_\gamma\sigma_z} (\sigmaa \sigmab + \sigmaa \sigmav{\bsy{\alpha}_1 \bsy{x'}} \sigmab \sigmav{\bsy{\alpha}_1 \bsy{x'}} )\right] \nonumber \\
&=& (-1)^N 8 \cos(2\theta_\gamma + \varphi_{\bsy{a}} - \varphi_{\bsy{\alpha}_1\bsy{x'}}) \cos(\varphi_{\bsy{b}} - \varphi_{\bsy{\alpha}_1\bsy{x'}}) \,,
\end{eqnarray}
where $\varphi_{\bsy{a}}$, $\varphi_{\bsy{b}}$, and $\varphi_{\bsy{\alpha}_1\bsy{x'}}$ are the polar angles of the vectors $\bsy{a}$, $\bsy{b}$, and $(\bsy{\alpha}_1 - \bsy{x'})$ respectively. 
We define the angles of the incoming ($\theta_b$) and the outgoing ($\theta_a$) segment of the trajectory $\gamma$ relative to the lead orientation (cf. Fig.\,\ref{Fig:loops}),
\begin{eqnarray}
\label{eq:def_out}
 \theta_a &=& \varphi_{\bsy{x}\bsy{\alpha}_N} - \varphi_{\bsy{a}} + \pi\,, \\
\label{eq:def_in}
 \theta_b &=& \varphi_{\bsy{\alpha}_1\bsy{x'}} -  \varphi_{\bsy{b}}\,,
\end{eqnarray}
and make use of the relation
\begin{equation}
\label{eq:total_open}
 2\theta_\gamma = N \pi + \varphi_{\bsy{\alpha}_1\bsy{x'}} - \varphi_{\bsy{x}\bsy{\alpha}_N} \mod 2\pi \,.
\end{equation}
so that we finally obtain
\begin{equation}
 \tr\left(\sigmaa \Ktilde_{\gamma}\sigmab  \Ktilde^\dagger_{\gamma}\right) = - 8 \cos(\theta_a)\cos(\theta_b)\,. 
\end{equation}
By using the same geometrical relations as in App.\,\ref{app:pspintrace_1} one shows that this holds also if reflections from zigzag edges are involved.
 
\subsection{Trace $Y_{\gamma,\gamma_\times}$ for the weak localization correction}
\label{app:pspintrace_4}

For the sum of reflection angles as defined in Eq.\,(\ref{eq:refang})
we have 
\begin{eqnarray}
 \theta_\gamma &=& \theta_{l_1} +\theta_{L}+ \theta_{l_2}\,, \\
 \theta_{\gamma_\times} &=& \theta_{l_1} -\theta_{L}+ \theta_{l_2}\,.
\end{eqnarray}
As in App.\,\ref{app:pspintrace_3}, we consider orbits with reflections from armchair and infinite mass edges to obtain for the sublattice part
\begin{eqnarray}
 &&\tr\left(\sigmaa \Ktilde^{\text{s}}_{\gamma}\sigmab  \Ktilde^{\text{s}\dagger}_{\gamma_\times}\right)\nonumber \\ &=&
 \tr\left[\sigmaa \sigma_z^N e^{i\theta_\gamma \sigma_z} (1+\sigmav{\bsy{\alpha}_1 \bsy{x'}}) 
 \sigmab (1+\sigmav{\bsy{\alpha}_1 \bsy{x'}}) e^{-i\theta_{\gamma_\times} \sigma_z} \sigma_z^N \right] \nonumber \\
&=& (-1)^N  \tr\left[e^{-2i(\theta_\gamma-\theta_L)\sigma_z} (\sigmaa \sigmab + \sigmaa \sigmav{\bsy{\alpha}_1 \bsy{x'}} \sigmab \sigmav{\bsy{\alpha}_1 \bsy{x'}} )\right] \nonumber \\
&=& - 4 \cos(\theta_a+2\theta_L )\cos(\theta_b) \nonumber \\ &=& (-1)^{N_L} 4 \cos(\theta_a)\cos(\theta_b) + \mathcal{O}(\varepsilon)\,,
\end{eqnarray}
where we used Eqs.\,(\ref{eq:def_out}-\ref{eq:total_open}) and the geometric relation
\begin{equation}
 2\theta_L = (N_{L} +1)\pi + \mathcal{O}(\varepsilon) \mod 2\pi\,.
\end{equation}
For the valley part we find that
\begin{eqnarray}
\hspace*{-0.6cm}
 \tr\left(\Ktilde^{\text{v}}_{\gamma} \Ktilde^{\text{v}\dagger}_{\gamma_\times}\right)
&=& \tr\left[\left(\Ktilde^{\text{v}}_L\right)^2\right] \nonumber \\
\hspace*{-0.6cm}&=& 
\left\{
\begin{array}{cl}
 (-1)^{N_{L} +1} \tr (\tau_0) & \text{for~odd~}N_{\text{ac},L} \\
 \tr [\exp(4iK \Lambda_L \tau_z)] & \text{for~even~}N_{\text{ac},L}
\end{array}
\right.  \nonumber \\
\hspace*{-0.6cm}&=& 2\left\{
\begin{array}{cl}
 (-1)^{N_{L} +1 } &  \text{for~odd~}N_{\text{ac},L}\,, \\
 \cos(4K \Lambda_L) &  \text{for~even~}N_{\text{ac},L}\,.
\end{array}
\right.
\end{eqnarray}
Thus for the whole pseudospin part, we have 
\begin{eqnarray}
 Y_{\gamma,\gamma_\times} &\approx& 
8\cos(\theta_a)\cos(\theta_b)  \\
&&\times \left\{
\begin{array}{cl}
 -1 & \text{for~odd~}N_{\text{ac},L}\,, \\
 (-1)^{N_{L}} \cos(4K \Lambda_L) & \text{for~even~}N_{\text{ac},L}\,,
\end{array}
\right. \nonumber
\end{eqnarray}
which can be generalized to include reflections from zigzag edges as in App.\,\ref{app:pspintrace_1} yielding Eq.\,(\ref{eq:Yggx})\,.

\subsection{Trace $X_{\gamma,\gamma',\rho,\rho'}$ for the universal conductance fluctuations}
\label{app:pspintrace_5}

As for the orbit pairs before, we consider quadruplets that hit only armchair and infinite mass type edges and consider the sublattice part first.
Then we have
\begin{eqnarray}
&& \tr\left(\sigmaa \Ktilde^{\text{s}}_{\gamma}\sigmaa  \Ktilde^{\text{s}\dagger}_{\gamma'}\right)  \\
&= &\tr\left[\sigmaa \sigma_z^{N_\gamma}e^{i\theta_\gamma\sigma_z}(1+\sigmav{\bsy{\alpha}_1\bsy{x}_a'})
\sigmaa (1+\sigmav{\bsy{\alpha}_1\bsy{x}_a'}) e^{-i\theta_{\gamma'}\sigma_z} \sigma_z^{N_{\gamma'}}\right] \nonumber \\ 
&=& (-1)^{N_\gamma} i^{N_\gamma+N_{\gamma'}} \nonumber \\ 
&\times& \tr\left[ e^{-i(\theta_\gamma+\theta_{\gamma'} + N_\gamma \pi/2 + N_{\gamma'} \pi/2)\sigma_z}
(1+[\sigmaa \sigmav{\bsy{\alpha}_1\bsy{x}_a'}]^2)\right] \nonumber \\
&=& (-1)^{N_\gamma} i^{N_\gamma+N_{\gamma'}} 4\, \cos(\theta_a')\nonumber \\ 
&\times& \cos(\theta_a' - \theta_\gamma - \theta_{\gamma'}- [N_\gamma + N_{\gamma'} ]\pi/2) \,. \nonumber
\end{eqnarray}
We make use of the identities
\begin{eqnarray}
\theta_\gamma + \theta_{\gamma'} &=& 2\theta_\gamma +\theta_q-\theta_r\,, \\
\theta_q-\theta_r &=& (N_{q,+} - N_{q,-} - N_{r,+} + N_{r,-}) \frac{\pi}{2} \nonumber \\
&-& w_q\pi+w_r\pi  + \mathcal{O}(\varepsilon_1, \varepsilon_2) \,, \\
 2\theta_\gamma &=& N_\gamma\pi + \varphi_{\bsy{\alpha}_1\bsy{x}_a'} - \varphi_{\bsy{x}_a\bsy{\alpha}_N} \mod 2\pi \nonumber \\
&=&  (N_\gamma+1)\pi + \theta_a' - \theta_a \mod 2\pi
\end{eqnarray}
and perform an analog calculation also for the second trace to obtain
\begin{eqnarray}
&&\tr\left(\sigmaa \Ktilde^{\text{s}}_{\gamma}\sigmaa  \Ktilde^{\text{s}\dagger}_{\gamma'}\right)
\tr\left(\sigmab \Ktilde^{\text{s}}_{\rho}\sigmab  \Ktilde^{\text{s}\dagger}_{\rho'}\right) \nonumber \\
&=&
16 \cos(\theta_a) \cos(\theta_a')\cos(\theta_b) \cos(\theta_b') \,,
\end{eqnarray}
if the orbit  $\rho$ ($\rho'$) traverses the part $q$ ($r$) in the same direction as $\gamma$ ($\gamma'$) [Fig.\,\ref{Fig:squids}\,a) and c)].
In the opposite case [Fig.\,\ref{Fig:squids}\,b) and d)], the result is 
\begin{eqnarray}
&&\tr\left(\sigmaa \Ktilde^{\text{s}}_{\gamma}\sigmaa  \Ktilde^{\text{s}\dagger}_{\gamma'}\right)
\tr\left(\sigmab \Ktilde^{\text{s}}_{\rho}\sigmab  \Ktilde^{\text{s}\dagger}_{\rho'}\right) \nonumber \\
&=&
(-1)^{N_L} 16 \cos(\theta_a) \cos(\theta_a')\cos(\theta_b) \cos(\theta_b') \,,
\end{eqnarray}
with $L= q + r^{-1}$.
Also for the valley part we have to distinguish the cases of parallel and antiparallel traversal of the loop by the orbits 
$\rho$ ($\rho'$) and $\gamma$ ($\gamma'$). We find that the valley part is zero for both cases, if $N_{\text{ac},L}$ is odd. For even $N_{\text{ac},L}$ we have 
for the parallel case
\begin{eqnarray}
&& \hspace*{-1.5cm} \tr\left(\Ktilde^{\text{v}}_{\gamma} \Ktilde^{\text{v}\dagger}_{\gamma'}\right)
 \tr\left(\Ktilde^{\text{v}}_{\rho} \Ktilde^{\text{v}\dagger}_{\rho'}\right) 
= \tr^2\left(\Ktilde^{\text{v}}_{q} \Ktilde^{\text{v}\dagger}_{r}\right) \nonumber \\
&=&
4 \cos^2(2K\Lambda_L + N_{\text{im},L} \,\pi/2)
\end{eqnarray}
and for the antiparallel case
\begin{eqnarray}
&& \hspace*{-0.5cm} \tr\left(\Ktilde^{\text{v}}_{\gamma} \Ktilde^{\text{v}\dagger}_{\gamma'}\right)
 \tr\left(\Ktilde^{\text{v}}_{\rho} \Ktilde^{\text{v}\dagger}_{\rho'}\right) 
= \tr\left(\Ktilde^{\text{v}}_{q} \Ktilde^{\text{v}\dagger}_{r}\right) \tr\left(\Ktilde^{\text{v}}_{q^{-1}} \Ktilde^{\text{v}\dagger}_{r^{-1}}\right) \nonumber \\
&=&
4 (-1)^{N_{\text{im},L}}\cos^2(2K\Lambda_L+ N_{\text{im},L}\,\pi/2)\,.
\end{eqnarray}
Again this result can be generalized for zigzag edges yielding Eq.\,(\ref{eq:X_ggrr})\,.

\section{Loop density and spectral form factor in the presence of a magnetic flux}
\label{app:FwM}
 
\subsection{Diagonal contribution to $F(t)$}
We consider a perpendicular magnetic field $B$, weak enough that we can neglect effects on the pseudospin or the classical orbits, i.\,e.\,we only take into account Aharonov-Bohm like phases
produced by the magnetic flux. An orbit $\gamma$ obtains a phase shift $\exp{( i \pi B A_\gamma/\Phi_0)}$, where $A_\gamma$ is the directed area enclosed by $\gamma$.
For pairs of identical orbits, the phase shifts cancel, but for the pairs $(\gamma, \gamma^{-1})$ the resulting phase difference is $2\pi B A_\gamma/\Phi_0$.
For chaotic systems the distribution of the $A_\gamma$ is known to be approximately Gaussian\cite{Berry1986}
\begin{equation}
 P_A(A_\gamma, T_\gamma) = \frac{1}{\sqrt{2\pi \alpha T_\gamma}} \exp\left(-\frac{A_\gamma^2  }{2 \alpha T_\gamma}\right)\,,
\end{equation}
with a system specific parameter $\alpha$.
Integrating the Aharonov-Bohm type phase shift over the enclosed areas with density $P_A$ results in an exponential suppression of the contribution of the corresponding orbit pairs:
\begin{equation}
\label{eq:Bdamping}
 \intl_{-\infty}^\infty dA_\gamma P_A(A_\gamma, T_\gamma) e^{2\pi i B A_\gamma/\Phi_0} = e^{-T_\gamma/\TB}
\end{equation}
with
\begin{equation}
\label{eq:TB}
 \TB = \frac{A^2}{2\pi^2\alpha} \frac{\Phi_0^2}{\Phi^2} = \zeta \frac{\Phi_0^2}{\Phi^2}\,.
\end{equation}
Since the pairs of identical orbits are not affected, we obtain for the diagonal contribution to the spectral form factor
\begin{equation}
  F_D(t) =  t \left(1+e^{-2 t T_{\text{H}}/T\subac} \right) \left(1+ e^{-t\Th/\TB}\right)\,,
\end{equation}
which turns into Eq.\,(\ref{eq:FD_gen}) for $\TB \rightarrow \infty$, i.\,e.\,zero magnetic field. For very strong ac scattering, $T\subac \rightarrow 0$, we recover the known formula for the GOE to GUE crossover Schr\"odinger billiard\cite{Bohigas1991}.

\subsection{Loop contribution to $F(t)$}
For the loop contribution the procedure to incorporate a flux is similar. Here we have always a phase difference $\exp{(i 2\pi B A_R/\Phi_0)}$ between the orbits $\gamma$ and $\gamma_\times$, so that the $T_R$ integration in Eq.\,(\ref{eq:modLoopdens}) is modified by an exponential such as in Eq.\,(\ref{eq:Bdamping}). However, one has to be more careful here when taking into account orbit pairs like $(\gamma,\gamma_\times^{-1})$, since they have the same propagation direction along $R$ and the opposite direction along $L$ in contrast to the pairs $(\gamma,\gamma_\times)$. Without magnetic field, this difference was irrelevant, but for finite $B$ we have to distinguish these contributions. The $T_R$ integral has to be modified by a factor
\begin{eqnarray}
&& \intl_{-\infty}^\infty dA_R P_A[A_R, T_R - \Tmin] e^{2\pi i B A_R/\Phi_0} \nonumber \\ &=& e^{-[T_R - \Tmin]/\TB}
 \end{eqnarray}
for the pairs $(\gamma,\gamma_\times)$ and by a factor
\begin{eqnarray}
 &&\intl_{-\infty}^\infty dA_L P_A[A_L,T_\gamma - T_R - \Tmin] e^{2\pi i B A_L/\Phi_0} \nonumber \\ &=& e^{-[T_\gamma - T_R - \Tmin]/\TB}
 \end{eqnarray}
for the pairs $(\gamma,\gamma_\times^{-1})$\cite{Turek2003}.
Analog to our discussion in \ref{ssec:F_loops}, we obtain for the loop contribution to the spectral form factor
\begin{eqnarray}
F_L(t) &\approx& -2t^2 \frac{\TB }{ \TB + T\subac/2} e^{-t \Th / \TB}\\
&& \times \left( 1- \left[ 1+ \frac{2\TB}{T\subac}\left( e^{t \Th / \TB} - 1  \right)  \right] e^{-2t \Th /T\subac} \right) \nonumber\,.
\end{eqnarray}
For zero magnetic field, $\TB \rightarrow \infty$, this formula turns into the former expression (\ref{eq:FL_gen}), while for finite magnetic field and very strong ac scattering, $T\subac \rightarrow 0$, we obtain again the known expression for the GOE to GUE crossover in a usual Schr\"odinger billiard\cite{Bohigas1991}:
\begin{equation}
 F_L(t) \approx -2t^2  e^{-t \Th / \TB}\,.
\end{equation}

\section{Connection of Kubo formula and scattering matrix formalism: Fisher-Lee relation and Landauer formula}
\label{app:FisherLee}
First we derive an equation that connects the Green's function of a scattering system to the elements of its scattering matrix via projection on the
lead channels, i.\,e.\,a graphene version of the Fisher-Lee relations \cite{Fisher1981}. To this end we define the wavefunction of a graphene lead
in channel $m$ as
\begin{equation}
 \Psi^{\pm}_m(\bsy{x}) = e^{\pm i |k_m|x}\, \psi^{\pm}_m(y)
\end{equation}
where $x$ and $y$ are the \textit{local} coordinates of the lead such that the positive $x$-axis points away from the scattering system.
The signs $\pm$ denote propagation in positive (i.\,e.\,outgoing modes) and negative (i.\,e.\,incoming modes) $x$-direction. 
While the transverse wavefunctions $\psi_m(y)$ of states with different $m$ are orthogonal in the case of the Schr\"odinger equation without magnetic field,
this is in general \textit{not the case} for graphene. However, the current operator is still diagonal in this subspace (as for the Schr\"odinger
equation with a magnetic field \cite{Baranger1989}), namely for wavefunctions normalized to unit flux we have
\begin{eqnarray}
\label{eq:current_ortho_1}
\int\limits_C dy \,\psi_{m'}^{\pm \dagger}(y)\, \sigma_{x}\, \psi_m^{\pm}(y) &=& \pm\frac{\delta_{mn}}{\vF}\,, \\
\label{eq:current_ortho_2}
\int\limits_C dy \,\psi_{m'}^{\mp \dagger}(y)\, \sigma_{x}\, \psi_m^{\pm}(y) &=& 0 \,,
\end{eqnarray}
where $C$ is the corresponding lead cross-section. To see this we consider two eigenmodes $m$ and $m'$ that fulfill the Dirac equation
\begin{eqnarray}
(\sigma_x\,  k_m +\sigma_y\,p_y/\hbar ) \ket{\psi_m} &=& \kE \ket{\psi_m}\,, \\
\bra{\psi_{m'}} (\sigma_x\, k_{m'} + \sigma_y\,p_y/\hbar ) &=&  \kE \bra{\psi_{m'}}\,.
\end{eqnarray}
We multiply the first equation with $\bra{\psi_m'}$ from the left and the second with $\ket{\psi_m}$ from the right and subtract the resulting equations
to get
\begin{equation}
 (k_{m'}-k_m) \bra{\psi_{m'}} \sigma_x \ket{\psi_m} =0\,.
\end{equation}
Then Eqs.\,(\ref{eq:current_ortho_1}) and (\ref{eq:current_ortho_2}) follow directly: If \mbox{$k_{m'}-k_m \neq 0$}, the matrix element has to vanish. Note that $m\neq m'$ does not necessarily imply $k_{m'}-k_m \neq 0$, e.\,g.\,for metallic armchair 
leads or if one considers spin degenerate modes. However in this case one can construct an orthogonal basis in the degenerate subspace so that the above still holds.

Now consider a scattering system that is connected to an arbitrary number of leads. If an electron enters from lead $\beta$ in the mode $m'$, the wavefunction in the \textit{asymptotic region}, i.\,e. far away from the scatterer, reads
\begin{equation}
\label{eq:Wfct_scatt}
 \Phi(\bsy{x}) = 
\left\{
\begin{array}{ll}
 \Psi_{m'}^-(\bsy{x}_{\beta}) + \sum\limits_m r_{mm'}^{(\beta)} \,\Psi_m^+(\bsy{x}_{\beta}) & \text{if~$\bsy{x} \in$ $\beta$}\,,\\
  \sum\limits_m t_{mm'}^{(\alpha\beta)} \,\Psi_m^+(\bsy{x}_{\alpha}) & \text{if~$\bsy{x}\in$ $\alpha \neq \beta$}\,,
 \end{array}
\right.
\end{equation}
where the sums run over all propagating modes $m$ in the corresponding lead. On the other hand the equation of motion for the retarded
Green's function gives
\begin{equation}
 \Phi(\bsy{x}) = \int\limits_\mathcal{V} d^2x'\, \underbrace{G(\bsy{x},\bsy{x'}) (\kE-i\bsy{\sigma}\cdot \overleftarrow{\bsy{\nabla}}_{\bsy{x'}}+i\eta)}_{=\delta(\bsy{x}-\bsy{x'})} \Phi(\bsy{x}')\,,
\end{equation}
with $\overleftarrow{\bsy{\nabla}}_{\bsy{x'}}$ acting to the left on the second argument of $G$.
Here $\mathcal{V}$ is a region in space that includes the scatterer completely and its boundary $\partial \mathcal{V}$ intersects the leads perpendicularly 
in the asymptotic region. Integration by parts then leads to
\begin{eqnarray}
\label{eq:Wfct_G}
 \Phi(\bsy{x}) 
= i\int\limits_{C_{\alpha}} dy'_{\alpha} \, G(\bsy{x},\bsy{x}_{\alpha}') 
 \,\sigma_{x_\alpha}\, \Psi_{m'}^-(\bsy{x}_{\alpha}')\,,
\end{eqnarray}
where $\bsy{n}_{\bsy{x'}}$ is the unit vector normal to the boundary $\mathcal{V}$ at the point $\bsy{x'}$ and $\sigma_{x_\alpha} = \bsy{\sigma} \cdot \hat{\bsy{x}}_\alpha$ . 
We used that in the asymptotic region the retarded Green's function contains only outgoing waves and can be expanded as\cite{Baranger1989}
\begin{equation}
\label{eq:Green_limit1}
 G(\bsy{x}_{\alpha},\bsy{x}_{\beta}') \stackrel{\bsy{x}_{\beta}'\rightarrow \infty}{\longrightarrow} \sum_m b_m(\bsy{x}_{\alpha}) \psi_m^{- \dagger}(\bsy{x}_{\beta}')\,.
\end{equation}
Further we employed the current orthogonality relations (\ref{eq:current_ortho_1}) and (\ref{eq:current_ortho_2}).
Now we multiply both Eq.\,(\ref{eq:Wfct_scatt}) and Eq.\,(\ref{eq:Wfct_G}) with the appropriate outgoing wavefunction in lead $\alpha$ and use again Eq.\,(\ref{eq:current_ortho_1}) to project out the transmission amplitude
\begin{eqnarray}
\label{eq:FisherLee}
 &&t_{mm'}^{(\alpha\beta)}  \\ &=& i\vF\int\limits_{C_{\alpha}} dy_{\alpha}\int\limits_{C_{\beta}} dy'_{\beta} \,
\Psi_m^{+\dagger}(\bsy{x}_\alpha) \, \sigma_{x_\alpha} \, G(\bsy{x}_{\alpha},\bsy{x'}_{\beta})\, \sigma_{x_\beta} \, \Psi_m^{-}(\bsy{x}_{\beta}') \nonumber
\end{eqnarray}
yielding the graphene version of the Fisher-Lee relations (see also Ref.\,\onlinecite{Carmier2011}).\\
We can now derive the Landauer formula for graphene that expresses the conductance as a sum over transmission probabilities of the
individual propagating modes in the leads. To this end 
we note that in analogy to Eq.\,(\ref{eq:Green_limit1}), the advanced Green's function
can be constructed solely out of incoming modes in the asymptotic region\cite{Baranger1989}. Thus, for the retarded Green's function it follows that
\begin{equation}
\label{eq:Green_limit2}
 G(\bsy{x}_{\alpha},\bsy{x}_{\beta}') \stackrel{\bsy{x}_{\alpha}\rightarrow \infty}{\longrightarrow} \sum_{m'}  
\psi_{m'}^{+}(\bsy{x}_{\alpha})\, d_{m'}^\dagger(\bsy{x}_{\beta}')\,,
\end{equation}
and in combination with Eq.\,(\ref{eq:Green_limit1}) we have
\begin{equation}
\label{eq:Green_limit3}
 G(\bsy{x}_{\alpha},\bsy{x}_{\beta}') \stackrel{\bsy{x}_{\alpha},\bsy{x}_{\beta}' \rightarrow \infty}{\longrightarrow} \sum_{mm'}  f_{mm'}
\psi_{m'}^{+}(\bsy{x}_{\alpha}) \psi_m^{- \dagger}(\bsy{x}_{\beta}')\,.
\end{equation}
Applying Eq.\,(\ref{eq:current_ortho_1}) twice gives for the coefficients \mbox{$f_{mm'} = i\vF\, t_{mm'}^{(\alpha\beta)}$}. We insert this expansion
into the linear response expression (\ref{eq:conductance}) for the conductance and finally use Eq.\,(\ref{eq:current_ortho_1}) again twice
to obtain the Landauer formula
\begin{equation}
\label{eq:landauer}
 g_{\alpha \beta} = \frac{e^2}{h} \sum_{mm'} |t_{mm'}^{\alpha \beta}|^2 = \frac{e^2}{h} \tr(tt^\dagger)\,, 
\end{equation}
where we defined $t_{mm'}=t_{mm'}^{(\alpha \beta)}$.

In reverse, we can obtain Eq.\,(\ref{eq:conductance}) from the Landauer formula (\ref{eq:landauer}) and the generalized Fisher-Lee relation (\ref{eq:FisherLee}) using
\begin{equation}
\label{eq:current_ortho_3}
\sum_{m'} \int\limits_C d\tilde{y} \,\psi_{m'}^{\pm}(y) \psi_{m'}^{\pm \dagger}(\tilde{y})\, \sigma_{x}\, \psi_m^{\pm}(\tilde{y}) = \pm\frac{\psi_{m}^{\pm}(y)}{\vF} \,,
\end{equation}
which follows directly from Eq.\,(\ref{eq:current_ortho_1}).
Inserting (\ref{eq:FisherLee}) into the Landauer formula (\ref{eq:landauer}) and using Eq.\,(\ref{eq:current_ortho_3}) twice we get then Eqs.\,(\ref{eq:conductance}) and (\ref{eq:conductivity}). 

In the same way, we further derive Eqs.\,(\ref{eq:ShotN_tttt}) and (\ref{eq:ShotN_sigma}), that are relevant in the context of shot noise.
Again with Eq.\,(\ref{eq:FisherLee}), we obtain 
\begin{eqnarray} 
 \tr(tt^\dagger tt^\dagger) &=& \sum_{\stackrel{\scriptstyle m,n \,\in \alpha}{o,p \,\in \beta}} t_{mo}\,t_{no}^*\,t_{np}\,t_{mp}^* 
 = \vF^4 \int\limits_{C_{\alpha/\beta}}\!\!\! dy_1 \ldots  dy_8 \nonumber \\
&\times& \psi_m^{+\dagger}(y_1)\,\sigma_{x_\alpha}\, G(\bsy{x_1},\bsy{x_2})\, \sigma_{x_\beta}\, \psi_o^{-}(y_2) \nonumber \\
&\times& \psi_o^{-\dagger}(y_3)\,\sigma_{x_\beta}\, G^\dagger(\bsy{x_4}, \bsy{x_3})\, \sigma_{x_\alpha}\, \psi_n^{+}(y_4) \nonumber \\
&\times& \psi_n^{+\dagger}(y_5)\,\sigma_{x_\alpha}\, G(\bsy{x_5},\bsy{x_6})\, \sigma_{x_\beta}\, \psi_p^{-}(y_6) \nonumber \\
&\times& \psi_p^{-\dagger}(y_7)\,\sigma_{x_\beta}\, G^\dagger(\bsy{x_8}, \bsy{x_7})\, \sigma_{x_\alpha}\, \psi_m^{+}(y_8)\,.
\end{eqnarray}
With the structure of the Green's function (\ref{eq:Green_limit3}) we then use Eq.\,(\ref{eq:current_ortho_3}) four times and
obtain after renaming the integration variables expressions (\ref{eq:ShotN_tttt}) and (\ref{eq:ShotN_sigma}).



\begin{thebibliography}{99}

\bibitem{PartI}
J. Wurm, K. Richter, and \.{I}. Adagideli, Phys. Rev. B {\bf 84}, 075468 (2011).

\bibitem{Berry1987}
M. Berry and R. Mondragon, Proc. R. Soc. Lond. A {\bf 412},  53  (1987).

\bibitem{Peres2010}
N.~M.~R. Peres, Rev. Mod. Phys. {\bf 82}, 2673 (2010)

\bibitem{Abergel2010}
D.~S.~L. Abergel, V. Apalkov, J. Berashevich, K. Ziegler, and T. Chakraborty, Adv. Phys. {\bf 59}, 261 (2010)

\bibitem{Das2011}
S. Das~Sarma, S.~Adam, E.~H. Hwang, and E. Rossi, Rev. Mod. Phys. {\bf 83}, 407 (2011)

\bibitem{Ponomarenko2008}
L.~A. Ponomarenko, F. Schedin, M.~I. Katsnelson, R. Yang, E.~W. Hill, K.~S. Novoselov, and A.~K. Geim, 
Science {\bf 320},  356  (2008).

\bibitem{Guttinger2008}
J.~G\"uttinger, C. Stampfer, S. Hellmüller, F. Molitor, T. Ihn, and K. Ensslin, 
Appl. Phys. Lett. {\bf 93},  212102  (2008).

\bibitem{Guttinger2010}
J. G\"uttinger, T. Frey, C. Stampfer, T. Ihn, and K. Ensslin, Phys. Rev. Lett. {\bf 105},  116801  (2010).

\bibitem{McCann2006}
E. McCann {\it et~al.}, Phys. Rev. Lett. {\bf 97},  146805  (2006).

\bibitem{Morozov2006}
S.~V. Morozov {\it et~al.}, Phys. Rev. Lett. {\bf 97},  016801  (2006).

\bibitem{Tikhonenko2008}
F.~V. Tikhonenko, D.~W. Horsell, R.~V. Gorbachev, and A.~K. Savchenko, Phys.
  Rev. Lett. {\bf 100},  056802  (2008).

\bibitem{Berezovsky2010}
J. Berezovsky and R.~M. Westervelt, Nanotech. {\bf 21},  274014  (2010).

\bibitem{Wu2007}
X. Wu, X. Li, Z. Song, C. Berger, and W.~A. de Heer, Phys. Rev. Lett. {\bf 98},  136801  (2007).

\bibitem{Mayorov2011}
A.~S. Mayorov, R.~V. Gorbachev, S.~V. Morozov, L. Britnell, R. Jalil, L.~A. Ponomarenko, P. Blake, K.~S. Novoselov, K. Watanabe, T. Taniguchi, and A.~K. Geim,
Nano Lett. {\bf 11},  2396–2399  (2011).

\bibitem{Eroms2009}
J. Eroms and D. Weiss, New J. Phys. {\bf 11},  095021  (2009).

\bibitem{Libisch2009}
F. Libisch, C. Stampfer, and J. Burgd\"orfer, Phys. Rev. B {\bf 79},  115423
  (2009).

\bibitem{Bardarson2009}
J.~H. Bardarson, M. Titov, and P.~W. Brouwer, Phys. Rev. Lett. {\bf 102},
  226803  (2009).

\bibitem{Yang2011}
R. Yang, L. Huang, Y.-C. Lai and C. Grebogi, Eur. Phys. Lett. {\bf 94}, 40004 (2011)

\bibitem{Wurm2009}
J. Wurm {\it et~al.}, Phys. Rev. Lett. {\bf 102},  056806  (2009).

\bibitem{Balian1970}
R. Balian and C. Bloch, Ann. Phys. {\bf 60},  401  (1970).

\bibitem{Hansson1983}
T.~H. Hansson and R.~L. Jaffe, Ann. Phys. {\bf 151}, 204 (1983).

\bibitem{Adagideli2002}
\.{I}. Adagideli and P.~M. Goldbart, Int. J. Mod. Phys. B {\bf 16},  1381  (2002).

\bibitem{Sieber2001}
M. Sieber and K. Richter, Phys.Scr. {\bf 128},  128  (2001).

\bibitem{Richter2002}
K. Richter and M. Sieber, Phys. Rev. Lett. {\bf 89},  206801  (2002).

\bibitem{Mathur1992}
H. Mathur and A.~D. Stone, Phys. Rev. Lett. {\bf 68},  2964  (1992).

\bibitem{Adagideli2010}
\.{I}. Adagideli {\it et~al.}, Phys. Rev. Lett. {\bf 105},  246807  (2010).

\bibitem{Zaitsev2005}
O. Zaitsev, D. Frustaglia and K. Richter, Phys. Rev. Lett. {\bf 94},
  026809  (2005).

\bibitem{Note1}
Also $\protect \mathcal {T}_z$ commutes with $H$, but for decoupled valleys it
  is identical to $\protect \mathcal {T}_0$ and for coupled valleys it is not
  conserved. Therefore we do not consider $\protect \mathcal {T}_z$ further.
  
\bibitem{Gnutzmann2004}
S. Gnutzmann and B. Seif, Phys. Rev. E {\bf 69}, 056219 (2004).

\bibitem{Haake2004}
F. Haake, {\em Quantum Signatures of Chaos} (Springer, Berlin, 2004). 

\bibitem{Berry1985}
M. Berry, Proc. R. Soc. Lond. A {\bf 400},  229  (1985).

\bibitem{Heusler2007}
S. Heusler, S. M\"uller, A. Altland, P. Braun, and F. Haake, Phys. Rev. Lett. {\bf 98},  044103  (2007).

\bibitem{Gutzwiller1990}
M. Gutzwiller, {\em Chaos in Classical and Quantum Mechanics} (Springer, New
  York, 1990), pp.\ 257--261.

\bibitem{Hannay1984}
J.~H. Hannay and A.~M. {Ozorio de Almeida}, J. Phys. A {\bf 17},  3429  (1984).

\bibitem{Note2}
Note that if \protect \emph {all} distances between the ac edges of the system
  fulfill the condition for metallic ac nanoribbons, this results in $K\Lambda
  _\gamma = 0 ~\protect \text {mod}~2\pi $ for all orbits and thus eventually $
  F_D(t) = 4 t$ follows. This is due to the fact that the symmetry $\protect
  \mathcal {T}_0$ remains intact in this case and gives rise to a valley-like
  subblock structure. While for chaotic billiards this is an unrealistic
  situation, the transport properties of metallic ac nanoribbons are affected
  by this valley-like substructure ( J. Wurm, M. Wimmer, \.{I}. Adagideli, and K. Richter, manuscript in preparation).

\bibitem{Legrand1991}
O. Legrand and D. Sornette, Phys. Rev. Lett. {\bf 66},  2172  (1991).

\bibitem{Baranger1989}
H.~U. Baranger and A.~D. Stone, Phys. Rev. B {\bf 40},  8169  (1989).

\bibitem{Nomura2007}
K. Nomura and A.~H. MacDonald, Phys. Rev. Lett. {\bf 98},  076602  (2007).

\bibitem{Crepieux2001}
A. Cr\'epieux and P. Bruno, Phys. Rev. B {\bf 64}, 014416 (2001)

\bibitem{Fisher1981}
D.~S. Fisher and P.~A. Lee, Phys. Rev. B {\bf 23},  6851  (1981).

\bibitem{Baranger1993}
H.~U. Baranger, R.~A. Jalabert, and A.~D. Stone, Phys. Rev. Lett. {\bf 70},
  3876  (1993).

\bibitem{Heusler2006}
S. Heusler, S. M\"uller, P. Braun, and F. Haake, Phys. Rev. Lett. {\bf 96},
  066804  (2006).

\bibitem{Bolte2007}
J. Bolte and D. Waltner, Phys. Rev. B {\bf 76},
  075330  (2007).

\bibitem{Baranger1994}
H.~U. Baranger and P.~A. Mello, Phys. Rev. Lett. {\bf 73},  142  (1994).

\bibitem{Jalabert1994}
R.~A. Jalabert, J.-L. Pichard, and C.~W.~J. Beenakker, Europhys. Lett.) {\bf 27},  255  (1994).

\bibitem{Sieber1999}
M. Sieber, J. Phys. A: Math. Gen. {\bf 32},  7679  (1999).

\bibitem{Note3}
Again a special situation occurs if all edges are of ac type and all
  distances fulfill the condition for metallic ac ribbons. Then $K\Lambda
  _L= 0~\protect \text {mod}~2\pi $ and hence $\left <Y_{\gamma ,\gamma
  _\times }\right >=0$ and the WL correction vanishes, indicating unitary
  symmetry.

\bibitem{Waltner2010}
D. Waltner and K. Richter, {\em Classical Correlations and Quantum Interference in Ballistic Conductors {\em in:}
Nonlinear Dynamics in Nanosystems},
  edited by G. Radons, B. Rumpf, and H.~G. Schuster (Wiley-VCH, New York,
  2010).

\bibitem{Zaslavsky1981}
G.~M. Zaslavsky, Phys. Rep. {\bf 80}, 157 (1981).

\bibitem{Aleiner1996}
I.~L. Aleiner and A.~I. Larkin, Phys. Rev. B {\bf 54}, 14423 (1996).

\bibitem{Adagideli2003}
\.{I}. Adagideli, Phys. Rev. B {\bf 68}, 233308 (2003).

\bibitem{Bunimovich1979}
L.~A. Bunimovich., Commun. Math. Phys. {\bf 65}, 295 (1979).

\bibitem{Baranger1993a}
H.~U. Baranger, R.~A. Jalabert, and A.~D. Stone, Chaos {\bf 3},  665  (1993).

\bibitem{Wimmer2009}
M. Wimmer and K. Richter, J. Comp. Phys. {\bf 228},  8548   (2009).

\bibitem{Kharitonov2008}
M.~Y. Kharitonov and K.~B. Efetov, Phys. Rev. B {\bf 78}, 033404  (2008).

\bibitem{Kechedzhi2008}
K. Kechedzhi, O. Kashuba, V.~I. Fal'ko, Phys. Rev. B {\bf 77}, 193403  (2008).

\bibitem{Lundeberg2009}
M.~B. Lundeberg and J.~A. Folk, Nat. Phys. {\bf 5}, 894 (2009).

\bibitem{Horsell2009}
D.~W. Horsell, A.~K. Savchenko, F.~V. Tikhonenko, K. Kechedzhi, I.~V. Lerner and V.~I. Fal'ko, Solid State Comm. {\bf 149}, 1041 (2009).

\bibitem{Berezovsky2010a}
J. Berezovsky, M.~F. Borunda, E.~J. Heller, and R.~M. Westervelt, Nanotech. {\bf 21}, 274013 (2010).

\bibitem{Brouwer2006}
P.~W. Brouwer and S. Rahav, Phys. Rev. B {\bf 74},  075322  (2006).

\bibitem{Beenakker1997}
C.~W.~J. Beenakker, Rev. Mod. Phys. {\bf 69},  731  (1997).

\bibitem{Tworzydlo2006}
J. Tworzyd\l{}o, B. Trauzettel, M. Titov, A. Rycerz, and C.~W.~J. Beenakker, Phys. Rev. Lett. {\bf 96},  246802 (2006).

\bibitem{Cheianov2006}
V.~V. Cheianov and V.~I. Fal'ko, Phys. Rev. B {\bf 74}, 041403  (2006).

\bibitem{Danneau2008}
R. Danneau, F. Wu, M.~F. Craciun, S. Russo, M.~Y. Tomi, J. Salmilehto, A.~F.~ Morpurgo, and P.~J. Hakonen, Phys. Rev. Lett. {\bf 100}, 196802 (2008).

\bibitem{DiCarlo2008}
L. DiCarlo, J.~R. Williams, Y. Zhang, D.T. McClure, and C.~M. Marcus, Phys. Rev. Lett. {\bf 100}, 156801 (2008).

\bibitem{Buttiker1990}
M. B\"uttiker, Phys. Rev. Lett. {\bf 65},  2901  (1990).

\bibitem{Lassl2003}
A. Lassl, Diploma thesis, Universit\"at Regensburg, (2003).

\bibitem{Muller2007}
S. M\"uller, S. Heusler, P. Braun, and F. Haake, New J. Phys. {\bf 9},  12
  (2007).

\bibitem{Schanz2003}
H. Schanz, M. Puhlmann, and T. Geisel, Phys. Rev. Lett. {\bf 91},  134101
  (2003).

\bibitem{Berry1986}
M.~V. Berry and M. Robnik, J. Phys. A {\bf 19},  649  (1986).

\bibitem{Bohigas1991}
O. Bohigas, {\em Les Houches 1989} (North Holland, Amsterdam, 1991)

\bibitem{Turek2003}
M. Turek and K. Richter, J. Phys. A {\bf 36},  L455  (2003).

\bibitem{Carmier2011}
P. Carmier, C. Lewenkopf, and D. Ullmo,  Phys. Rev. B {\bf 84}, 195428 (2011)

\end{thebibliography}
\end{document}